\documentclass[12pt]{article}

\usepackage{graphics}
\usepackage{epsfig}
\input epsf

\title{Production of electroweak gauge bosons in off-shell \\gluon-gluon fusion}

\author{S.P.~Baranov$^a$, A.V.~Lipatov$^b$, N.P.~Zotov$^b$}

\begin{document}

\maketitle

\begin{center}

{\it $^a$\,P.N.~Lebedev Physics Institute,\\ 
119991 Moscow, Russia\/}\\[3mm]

{\it $^b$\,D.V.~Skobeltsyn Institute of Nuclear Physics,\\ 
M.V. Lomonosov Moscow State University,
\\119991 Moscow, Russia\/}\\[3mm]

\end{center}

\vspace{0.5cm}

\begin{center}

{\bf Abstract }

\end{center}

We study the production of electroweak gauge bosons at 
high energies in the framework of $k_T$-factorization QCD
approach. 
The amplitude for production of a single $W^\pm$ or $Z^0$ boson associated with 
quark pair in the fusion of two off-shell gluons is calculated. 
Contributions from the valence quarks are calculated
using the quark-gluon interaction and quark-antiquark 
annihilation QCD subprocesses.
The total and differential cross sections (as a function
of the transverse momentum and rapidity) are
presented and the ratio of cross sections for $W^\pm$ and $Z^0$ boson
production is investigated. 
The conservative error analysis is performed.
In the numerical calculations two different 
sets of unintegrated gluon distributions in the proton
are used: the one
obtained from Ciafaloni-Catani-Fiorani-Marchesini evolution equation
and the other from Kimber-Martin-Ryskin prescription.
Theoretical results are compared with 
experimental data taken by the D$\oslash$ and CDF collaborations 
at the Tevatron. We demonstrate the importance of 
the quark component in parton evolution in description
of the experimental data. This component is very significant also
at the LHC energies.

\vspace{1cm}

\noindent
PACS number(s): 12.38.Bx, 12.15.Ji

\vspace{0.5cm}

\section{Introduction} \indent 

The theoretical and experimental studying
the vector ($W^\pm$ and $Z^0$) boson production at high energies provide
information about the nature of both the underlying electroweak
interaction and the effects of Quantum Chromodynamics (QCD).
In many respects these processes have become
one of most important "standard candles" in experimental
high energy physics~[1--9]. At the Tevatron, measurements of $W^\pm$ and $Z^0$
inclusive cross sections are routinely used to validate
detector and trigger perfomance and stability. 
Data from gauge boson production also provide
bounds on parametrizations used to describe the non-perturbative
regime of QCD processes.
At the LHC, such measurements can serve as a useful tool to determine
the integrated luminosity and can also be used
to normalize measurements of other production
cross sections (for example, cross section of $W + n$-jets or 
diboson production). 
Additionally, studying of inclusive vector boson 
production is necessary starting point for investigations of 
Higgs or top quark production where many signatures can include these bosons.

At leading order (LO) of QCD, $W^\pm$ and $Z^0$ bosons are produced
via quark-antiquark annihilation. Beyond the LO Born process,
vector boson can also be produced by $q + g$ interactions,
so both the quark and gluon distribution functions of the proton
play an important role. Theoretical calculations of the 
$W^\pm$ and $Z^0$ production cross sections have been 
carried out at next-to-leading order (NLO) and next-to-next-to-leading
order (NNLO)~[10--14] of QCD. The NLO cross section is $\sim 25$\%
larger than the Born-level cross section, and the NNLO cross
section is an additional $\sim 3$\% higher. 
However, these perturbative calculations are reliable at high $p_T$ only
since diverge in the small $p_T \ll m$ region with terms
proportional to $\ln m/p_T$ (appearing due to soft
and collinear gluon emission). Therefore, the soft gluon
resummation technique~[15--19] should be used to make QCD predictions at low $p_T$.
The traditional calculations combine fixed-order
perturbation theory with analytic resummation and some matching criterion.
The analytic resummation can be performed either
in the transverse momentum space~[20] or in the Fourier
conjugate impact parameter space~[21].
Differences between the two formalisms are discussed
in~[22].

An alternative description can be provided by the
$k_T$-factorization approach of QCD~[23, 24].
This approach is based on the familiar
Balitsky-Fadin-Kuraev-Lipatov (BFKL)~[25] 
or Catani-Ciafaloni-Fiorani-Marchesini (CCFM)~[26] gluon evolution 
equations and takes into account
the large logarithmic terms proportional to $\ln 1/x$.
This contrasts with the usual 
Dokshitzer-Gribov-Lipatov-Altarelli-Parisi 
(DGLAP)~[27] strategy where only the large 
logarithmic terms proportional to $\ln \mu^2$ are taken into account.
The basic dynamical quantity of the $k_T$-factorization approach is 
the unintegrated (i.e., ${\mathbf k}_T$-dependent) parton distribution 
$f_a(x,{\mathbf k}_T^2,\mu^2)$ which determines the probability to find a 
type $a$ parton carrying the longitudinal momentum fraction $x$ and the 
transverse momentum ${\mathbf k}_T$ at the probing scale $\mu^2$.
In this approach, since each incoming parton carries its own nonzero
transverse momentum, the Born-level subprocess 
$q + \bar q^\prime \to W^\pm/Z^0$ already generate the $p_T$ distribution
of produced vector boson.
Similar to DGLAP, to calculate the cross sections of any 
physical process the unintegrated parton density 
$f_a(x,{\mathbf k}_T^2,\mu^2)$ 
has to be convoluted~[23, 24] with the relevant partonic cross section 
which has to be taken off mass shell (${\mathbf k}_T$-dependent). 
The soft gluon resummation formulas 
are the result of the approximate treatment of 
the solutions of the CCFM evolution equation~[28].
Other important properties of the 
$k_T$-factorization formalism are the additional contribution to the cross 
sections due to the integration over the ${\mathbf k}_T^2$ region above $\mu^2$
and the broadening of the transverse momentum distributions due to extra 
transverse momentum of the colliding partons\footnote{For an introduction
to $k_T$-factorization, see, for example, review~[29].}.

The $k_T$-factorization formalism has been already 
applied~[30] to calculate transverse momentum distribution of the 
inclusive $W^\pm$ and $Z^0$ production at Tevatron. 
The calculations~[30] were based on the usual 
(on-mass shell) matrix elements of the quark-antiquark 
annihilation subprocess $q + \bar q^\prime \to W^\pm/Z^0$
which embedded in precise off-shell kinematics.
However, an important component of the calculations~[30] is the 
unintegrated quark distribution in a proton. At present these distributions
are only available in the framework of the Kimber-Martin-Ryskin (KMR) approach~[31] 
since there are some theoretical difficulties 
in obtaining the quark densities immediately from CCFM or BFKL 
equations\footnote{Unintegrated quark density was considered recently in~[32].}
(see, for example, review~[29] for more details).
As a result the dependence of the calculated cross sections 
on the non-collinear evolution scheme has not been investigated.
This dependence in general can be significant and it is a special 
subject of study in the $k_T$-factorization formalism. 
Therefore in the present paper we will try a different and more systematic way.
Instead of using the unintegrated quark distributions and 
the corresponding quark-antiquark annihilation cross
section we calculate off-shell matrix element of 
the $g^* + g^* \to W^\pm/Z^0 + q \bar q^\prime$ subprocess
and then operate in terms of the unintegrated gluon 
densities only. In this scenario, at the price of considering the 
$2 \to 3$ rather than $2 \to 1$ matrix 
elements, the problem of unknown unintegrated quark
distributions will reduced to the problem of gluon distributions.
However, since the gluons are only responsible for the appearance of the sea but 
not valence quarks, the contribution from the valence quarks should be 
calculated separately. Having in mind that the valence quarks are only 
important at large $x$, where the traditional DGLAP evolution is accurate and 
reliable, this contribution can be taken into account within the usual collinear 
scheme based on the $q + g^* \to W^\pm/Z^0 + q^\prime$ and 
$q + \bar q^\prime \to W^\pm/Z^0$ matrix elements
convoluted with the on-shell valence quark and/or off-shell gluon 
densities\footnote{To avoid the double counting we have not considered here
$q + \bar q^\prime \to W^\pm/Z^0 + g$ subprocess.}.
Thus, the proposed way enables us with making comparisons 
between the different parton 
evolution schemes and parametrizations of parton 
densities\footnote{The similar scenario has been applied recently
to the prompt photon hadroproduction at Tevatron~[33].}.

We should mention, of course, that this idea can only
work well if the sea quarks appear from the last step
of the gluon evolution --- then we can absorb this
last step of the gluon ladder into hard matrix element.
However, this method does not apply to the quarks
coming from the earlier steps of the evolution
(i.e., from the second-to-last, third-to-last and other gluon
splittings). But it is not evident in advance, whether
the last gluon splitting dominates or not. The goal of our
study is to clarify this point.

The outline of our paper is following. In Section~2 we 
recall shortly the basic formulas of the $k_T$-factorization approach with a brief 
review of calculation steps and the unintegrated 
parton densities used. We will concentrate mainly
on the $g^* + g^* \to W^\pm/Z^0 + q \bar q^\prime$
subprocess. The evaluation of $q + g^* \to W^\pm/Z^0 + q^\prime$ and 
$q + \bar q^\prime \to W^\pm/Z^0$ contributions is
straightforward and, for the reader's convenience, we only collect 
the main relevant formulas in Appendix.
In Section~3 we present the numerical results
of our calculations. 
The central point is discussing the role
of each contribution mentioned above to the cross section.
Special attention is put on the transverse momentum
distributions of the $W^\pm$ and $Z^0$ boson 
measured by the D$\oslash$~[5, 8, 9] and CDF~[4] collaborations.
Section~4 contains our conclusions.

\section{Theoretical framework} \indent

As the off-shell gluon-gluon fusion $g^* + g^* \to W^\pm/Z^0 + q \bar q^\prime$
is calculated for the first time in the literature,
we find it reasonable to explain it in more detail.

\subsection{Kinematics} \indent 

We start from the kinematics (see Fig.~1). 
Let $p^{(1)}$ and $p^{(2)}$ be the four-momenta of the incoming protons and 
$p$ the four-momentum of the produced $W^\pm$/$Z^0$ boson.
The initial off-shell gluons have the four-momenta
$k_1$ and $k_2$ and the final quark $q$ and antiquark $\bar q^\prime$ have the 
four-momenta $p_1$ and $p_2$ and the masses $m_1$ and $m_2$, respectively.
In the $p \bar p$ center-of-mass frame we can write
$$
  p^{(1)} = {\sqrt s\over 2} (1,0,0,1),\quad p^{(2)} = {\sqrt s\over 2} (1,0,0,-1), \eqno(1)
$$

\noindent
where $\sqrt s$ is the total energy of the process 
under consideration and we neglect the masses of the incoming protons.
The initial gluon four-momenta in the high energy limit can be written as
$$
  k_1 = x_1 p^{(1)} + k_{1T},\quad k_2 = x_2 p^{(2)} + k_{2T}, \eqno(2)
$$

\noindent 
where $k_{1T}$ and $k_{2T}$ are the transverse four-momenta.
It is important that ${\mathbf k}_{1T}^2 = - k_{1T}^2 \neq 0$ and
${\mathbf k}_{2T}^2 = - k_{2T}^2 \neq 0$. From the conservation laws 
we can obtain the following relations:
$$
  {\mathbf k}_{1T} + {\mathbf k}_{2T} = {\mathbf p}_{1T} + {\mathbf p}_{2T} + {\mathbf p}_{T},
$$
$$
  x_1 \sqrt s = m_{1T} e^{y_1} + m_{2T} e^{y_2} + m_T e^y, \eqno(3)
$$
$$
  x_2 \sqrt s = m_{1T} e^{-y_1} + m_{2T} e^{-y_2} + m_T e^{-y},
$$

\noindent 
where $p_T$, $m_T$ and $y$ are the transverse momentum, 
transverse mass and center-of-mass rapidity of produced $W^\pm$/$Z^0$ boson, 
$p_{1T}$ and $p_{2T}$ are the transverse momenta of final quark
$q$ and antiquark $\bar q^\prime$, $y_1$, $y_2$, $m_{1T}$ and $m_{2T}$ 
are their rapidities and 
transverse masses, i.e. $m_{iT}^2 = m_i^2 + {\mathbf p}_{iT}^2$.

\subsection{Off-shell amplitude of the $g^* + g^* \to W^\pm/Z^0 + q \bar q^\prime$ subprocess} \indent 

There are eight Feynman diagrams (see Fig.~2) which describe the partonic
subprocess $g^* + g^* \to W^\pm/Z^0 + q \bar q^\prime$ at $\alpha \alpha_s^2$ order.
Let $\epsilon_1$, $\epsilon_2$ and $\epsilon$ be the initial gluon and produced gauge boson 
polarization vectors, respectively, and $a$ and $b$ the eight-fold color indices of the off-shell
gluons.
Then the relevant matrix element can be presented as follows:
$$
  {\cal M}_1 = g^2 \, \bar u (p_1) \, t^a \gamma^\mu \epsilon_\mu {\hat p_1 - \hat k_1 + m_1\over m_1^2 - (p_1 - k_1)^2} T_{W,Z}^\lambda \, \epsilon_\lambda {\hat k_2 - \hat p_2 + m_2\over m_2^2 - (k_2 - p_2)^2} t^b \gamma^\nu \epsilon_\nu \, u(p_2), \eqno(4)
$$
$$
  {\cal M}_2 = g^2 \, \bar u (p_1) \, t^b \gamma^\nu \epsilon_\nu {\hat p_1 - \hat k_2 + m_1\over m_1^2 - (p_1 - k_2)^2} T_{W,Z}^\lambda \, \epsilon_\lambda {\hat k_1 - \hat p_2 + m_2\over m_2^2 - (k_1 - p_2)^2} t^a \gamma^\mu \epsilon_\mu \, u(p_2), \eqno(5)
$$
$$
  {\cal M}_3 = g^2 \, \bar u (p_1) \, t^a \gamma^\mu \epsilon_\mu {\hat p_1 - \hat k_1 + m_1\over m_1^2 - (p_1 - k_1)^2}\, t^b \gamma^\nu \epsilon_\nu { - \hat p_2 - \hat p + m_1\over m_1^2 - ( - p_2 - p)^2} T_{W,Z}^\lambda \, \epsilon_\lambda \, u(p_2), \eqno(6)
$$
$$
  {\cal M}_4 = g^2 \, \bar u (p_1) \, t^b \gamma^\nu \epsilon_\nu {\hat p_1 - \hat k_2 + m_1\over m_1^2 - (p_1 - k_2)^2}\, t^a \gamma^\mu \epsilon_\mu { - \hat p_2 - \hat p + m_1\over m_1^2 - ( - p_2 - p)^2} T_{W,Z}^\lambda \, \epsilon_\lambda \, u(p_2), \eqno(7)
$$
$$
  {\cal M}_5 = g^2 \, \bar u (p_1) \, T_{W,Z}^\lambda \, \epsilon_\lambda {\hat p_1 + \hat p + m_2\over m_2^2 - (p_1 + p)^2}\, t^b \gamma^\nu \epsilon_\nu { \hat k_1 - \hat p_2 + m_2\over m_2^2 - (k_1 - p_2)^2} t^a \gamma^\mu \epsilon_\mu \, u(p_2), \eqno(8)
$$
$$
  {\cal M}_6 = g^2 \, \bar u (p_1) \, T_{W,Z}^\lambda \, \epsilon_\lambda {\hat p_1 + \hat p + m_2\over m_2^2 - (p_1 + p)^2}\, t^a \gamma^\mu \epsilon_\mu { \hat k_2 - \hat p_2 + m_2\over m_2^2 - (k_2 - p_2)^2} t^b \gamma^\nu \epsilon_\nu \, u(p_2), \eqno(9)
$$
$$
  \displaystyle {\cal M}_7 = g^2 \, \bar u (p_1) \, \gamma^\rho C^{\mu \nu \rho}(k_1,k_2,- k_1 - k_2){\epsilon_\mu \epsilon_\nu \over (k_1 + k_2)^2} f^{abc} t^c \times \atop 
  \displaystyle \times { - \hat p_2 - \hat p + m_1\over m_1^2 - ( - p_2 - p)^2}\, T_{W,Z}^\lambda \, \epsilon_\lambda \, u(p_2), \eqno(10)
$$
$$
  \displaystyle {\cal M}_8 = g^2 \, \bar u (p_1) \, T_{W,Z}^\lambda \, \epsilon_\lambda {\hat p_1 + \hat p + m_2\over m_2^2 - (p_1 + p)^2} \times \atop 
  \displaystyle \times \gamma^\rho C^{\mu \nu \rho}(k_1,k_2,- k_1 - k_2) {\epsilon_\mu \epsilon_\nu \over (k_1 + k_2)^2} f^{abc} t^c \, u(p_2). \eqno(11)
$$

\vspace{0.2cm}

\noindent
In the above expressions $C^{\mu \nu \rho}(k,p,q)$ and $T_{W,Z}^\lambda$ are related to the standard QCD
three-gluon coupling and the $W^\pm$/$Z^0$-fermion vertexes:
$$
  C^{\mu \nu \rho}(k,p,q) = g^{\mu \nu} (p - k)^\rho + g^{\nu \rho} (q - p)^\mu + g^{\rho \mu} (k - q)^\nu, \eqno(12)
$$
$$
  T^\lambda_W = {e\over 2 \sqrt 2 \sin \theta_W} \gamma^\lambda (1 - \gamma^5) V_{qq^\prime}, \eqno(13)
$$
$$
  T^\lambda_Z = {e\over \sin 2 \theta_W} \gamma^\lambda \left[I_3^{(q)}(1 - \gamma^5) - 2 e_q \sin^2 \theta_W\right], \eqno(14)
$$

\noindent
where $I_3^{(q)}$ and $e_q$ are the weak isospin and the fractional electric charge 
(in the positron charge $e$ units) of final-state quark $q$,  
$\theta_W$ is the Weinberg mixing angle and $V_{qq^\prime}$ is the
Cabibbo-Kobayashi-Maskawa (CKM) matrix element. Of course, in
the case of $Z^0$ production $m_1$ equals $m_2$.
The summation on the $W^\pm$/$Z^0$ polarization is carried out by the
covariant formula
$$
  \sum \epsilon^\mu (p) \, \epsilon^{* \, \nu} (p) = - g^{\mu \nu} + {p^\mu p^\nu\over m^2}. \eqno(15)
$$

\noindent
In the case of the initial off-shell gluon we use the BFKL prescription~[23, 24]:
$$
  \sum \epsilon^\mu (k_i) \, \epsilon^{* \, \nu} (k_i) = {k_{iT}^\mu k_{iT}^\nu \over {\mathbf k}_{iT}^2}. \eqno(16)
$$

\noindent
This formula converges to the usual expression 
$\sum \epsilon^\mu \epsilon^{* \, \nu} = -g^{\mu \nu}$ 
after azimuthal angle averaging
in the $k_T \to 0$ limit. 
The evaluation of the traces in~(4) --- (11) was done using the algebraic 
manipulation system \textsc{Form}~[34]. 
We would like to mention here that the usual method 
of squaring of~(4) --- (11) results in enormously long
output. This technical problem was solved by applying the
method of orthogonal amplitudes~[35].

The gauge invariance of the matrix element is a
subject of special attention in the $k_T$-factorization approach. Strictly speaking,
the diagrams shown in Fig.~2 are insufficient and have to be accompanied
with the graphs involving direct gluon exchange between the protons
(these protons are not shown in Fig.~2). These graphs are 
necessary to maintain the gauge invariance.
However, they violate the factorization since they cannot be represented
as a convolution of the gluon-gluon fusion matrix element with unintegrated gluon density.
The solution pointed out in~[24] refers to the fact that, within the 
particular gauge~(16), the contribution from these unfactorizable diagrams
vanish, and one has to only take into account the graphs depicted in Fig.~2.
We have successfully tested the gauge invariance of the matrix 
element~(4) --- (11) numerically\footnote{At the
preliminary stage of the work we have made a cross-check 
of the matrix elements which have been 
calculated independently by M.~Deak and F.~Schwennsen.}.

\subsection{Cross section for the inclusive $W^\pm$/$Z^0$ production} \indent 

According to the $k_T$-factorization theorem, the 
inclusive $W^\pm$/$Z^0$ production cross section 
via two off-shell gluon fusion 
can be written as a convolution
$$
  \displaystyle \sigma (p + \bar p \to W^\pm/Z^0 + X) = \sum_{q} \int {dx_1\over x_1} f_g(x_1,{\mathbf k}_{1 T}^2,\mu^2) d{\mathbf k}_{1 T}^2 {d\phi_1\over 2\pi} \times \atop 
  \displaystyle \times \int {dx_2\over x_2} f_g(x_2,{\mathbf k}_{2 T}^2,\mu^2) d{\mathbf k}_{2 T}^2 {d\phi_2\over 2\pi} d{\hat \sigma} (g^* + g^* \to W^\pm/Z^0 + q \bar q^\prime), \eqno(17)
$$

\noindent 
where $\hat \sigma(g^* + g^* \to W^\pm/Z^0 + q \bar q^\prime)$ is the partonic cross section, 
$f_g(x,{\mathbf k}_{T}^2,\mu^2)$ is the unintegrated gluon distribution in a proton 
and $\phi_1$ and $\phi_2$ are the azimuthal angles of the incoming gluons.
The multiparticle phase space $\Pi d^3 p_i / 2 E_i \delta^{(4)} (\sum p^{\rm in} - \sum p^{\rm out} )$
is parametrized in terms of transverse momenta, rapidities and azimuthal angles:
$$
  { d^3 p_i \over 2 E_i} = {\pi \over 2} \, d {\mathbf p}_{iT}^2 \, dy_i \, { d \phi_i \over 2 \pi}. \eqno(18)
$$

\noindent
Using the expressions~(17) and~(18) we obtain the master formula:
$$
  \displaystyle \sigma(p + \bar p \to W^\pm/Z^0 + X) = \sum_{q} \int {1\over 256\pi^3 (x_1 x_2 s)^2} |\bar {\cal M}(g^* + g^* \to W^\pm/Z^0 + q \bar q^\prime)|^2 \times \atop 
  \displaystyle \times f_g(x_1,{\mathbf k}_{1 T}^2,\mu^2) f_g(x_2,{\mathbf k}_{2 T}^2,\mu^2) d{\mathbf k}_{1 T}^2 d{\mathbf k}_{2 T}^2 d{\mathbf p}_{1 T}^2 {\mathbf p}_{2 T}^2 dy dy_1 dy_2 {d\phi_1\over 2\pi} {d\phi_2\over 2\pi} {d\psi_1\over 2\pi} {d\psi_2\over 2\pi}, \eqno(19)
$$

\noindent
where $|\bar {\cal M}(g^* + g^* \to W^\pm/Z^0 + q \bar q^\prime)|^2$ is the off-mass shell 
matrix element squared and averaged over initial gluon 
polarizations and colors, $\psi_1$ and $\psi_2$ are the 
azimuthal angles of the final state quark and antiquark, respectively.
We would like to point out again that $|\bar {\cal M}(g^* + g^* \to W^\pm/Z^0 + q \bar q^\prime)|^2$
strongly depends on the nonzero 
transverse momenta ${\mathbf k}_{1 T}^2$ and ${\mathbf k}_{2 T}^2$.
If we average the expression~(19) over $\phi_{1}$ and $\phi_{2}$ 
and take the limit ${\mathbf k}_{1 T}^2 \to 0$ and ${\mathbf k}_{2 T}^2 \to 0$,
then we recover the expression for the $W^\pm$/$Z^0$ production cross section in the  
collinear $\alpha \alpha_s^2$ approximation.

The multidimensional integration in~(19) has been performed
by means of the Monte Carlo technique, using the routine 
\textsc{Vegas}~[36]. The full C$++$ code is available from the 
authors upon request\footnote{lipatov@theory.sinp.msu.ru}.

\subsection{The KMR unintegrated parton distributions} \indent 

In the present paper we have tried two different sets of 
unintegrated parton densities in a proton. First of them
is the Kimber-Martin-Ryskin set.

The KMR approach~[31] is the formalism to construct
parton distributions $f_a(x,{\mathbf k}_T^2,\mu^2)$ unintegrated over the parton 
transverse momenta ${\mathbf k}_T^2$ from the known conventional parton
distributions $xa(x,\mu^2)$, where $a = g$ or $a = q$. This formalism 
is valid for a proton as well as a photon and
can embody both DGLAP and BFKL contributions. It also accounts for 
the angular ordering which comes from coherence effects in gluon emission.
The key observation here is that the $\mu$ dependence of the unintegrated 
parton distributions $f_a(x,{\mathbf k}_T^2,\mu^2)$ enters at the last step
of the evolution, and therefore single scale evolution equations (pure DGLAP)
can be used up to this step. In this approximation, the unintegrated quark and 
gluon distributions are given by the expressions~[31]
$$
  \displaystyle f_q(x,{\mathbf k}_T^2,\mu^2) = T_q({\mathbf k}_T^2,\mu^2) {\alpha_s({\mathbf k}_T^2)\over 2\pi} \times \atop {
  \displaystyle \times \int\limits_x^1 dz \left[P_{qq}(z) {x\over z} q\left({x\over z},{\mathbf k}_T^2\right) \Theta\left(\Delta - z\right) + P_{qg}(z) {x\over z} g\left({x\over z},{\mathbf k}_T^2\right) \right],} \eqno (20)
$$
$$
  \displaystyle f_g(x,{\mathbf k}_T^2,\mu^2) = T_g({\mathbf k}_T^2,\mu^2) {\alpha_s({\mathbf k}_T^2)\over 2\pi} \times \atop {
  \displaystyle \times \int\limits_x^1 dz \left[\sum_q P_{gq}(z) {x\over z} q\left({x\over z},{\mathbf k}_T^2\right) + P_{gg}(z) {x\over z} g\left({x\over z},{\mathbf k}_T^2\right)\Theta\left(\Delta - z\right) \right],} \eqno (21)
$$

\noindent
where $P_{ab}(z)$ are the usual unregulated leading order DGLAP splitting 
functions, and $q(x,\mu^2)$ and $g(x,\mu^2)$ are the conventional quark 
and gluon densities\footnote{Numerically, we have used the 
standard GRV~(LO) parametrizations~[37].}. The theta functions which appear 
in~(20) and~(21) imply 
the angular-ordering constraint $\Delta = \mu/(\mu + |{\mathbf k}_T|)$ 
specifically to the last evolution step to regulate the soft gluon
singularities. For other evolution steps, the strong ordering in 
transverse momentum within the DGLAP equations automatically 
ensures angular ordering. It is important that the parton 
distributions $f_a(x,{\mathbf k}_T^2,\mu^2)$ extended now into 
the ${\mathbf k}_T^2 > \mu^2$ region. This fact is in the clear contrast with the 
usual DGLAP evolution\footnote{We would like to note that 
cut-off $\Delta$ can be taken $\Delta = |{\mathbf k}_T|/\mu$ also~[31]. 
In this case the unintegrated parton distributions given by (20) --- (21) 
vanish for ${\mathbf k}_T^2 > \mu^2$ in accordance with 
the DGLAP strong ordering in ${\mathbf k}_T^2$.}.

The virtual (loop) contributions may be resummed 
to all orders by the quark and gluon Sudakov form factors
$$
  \ln T_q({\mathbf k}_T^2,\mu^2) = - \int\limits_{{\mathbf k}_T^2}^{\mu^2} {d {\mathbf p}_T^2\over {\mathbf p}_T^2} {\alpha_s({\mathbf p}_T^2)\over 2\pi} \int\limits_0^{z_{\rm max}} dz P_{qq}(z), \eqno (22)
$$
$$
  \ln T_g({\mathbf k}_T^2,\mu^2) = - \int\limits_{{\mathbf k}_T^2}^{\mu^2} {d {\mathbf p}_T^2\over {\mathbf p}_T^2} {\alpha_s({\mathbf p}_T^2)\over 2\pi} \left[ n_f \int\limits_0^1 dz P_{qg}(z) + \int\limits_{z_{\rm min}}^{z_{\rm max}} dz z P_{gg}(z) \right], \eqno (23)
$$

\noindent
where $z_{\rm max} = 1 - z_{\rm min} = {\mu/({\mu + |{\mathbf p}_T|}})$.
The form factors $T_a({\mathbf k}_T^2,\mu^2)$ give the probability of 
evolving from a scale ${\mathbf k}_T^2$ to a scale $\mu^2$ without 
parton emission. In according with~(22) and~(23)
$T_a({\mathbf k}_T^2,\mu^2) = 1$ in the ${\mathbf k}_T^2 > \mu^2$ region.

Note that such definition of the $f_a(x,{\mathbf k}_T^2,\mu^2)$ is 
correct for ${\mathbf k}_T^2 > \mu_0^2$ only, where 
$\mu_0 \sim 1$ GeV is the minimum scale for which DGLAP evolution of 
the collinear parton densities is valid. Everywhere in our numerical 
calculations we set the starting scale $\mu_0$ to be equal $\mu_0 = 1$ GeV.
Since the starting point of this derivation is the leading order 
DGLAP equations, the unintegrated parton distributions must satisfy
the normalisation condition
$$
  a(x,\mu^2) = \int\limits_0^{\mu^2} f_a(x,{\mathbf k}_T^2,\mu^2) d{\mathbf k}_T^2. \eqno(24)
$$

\noindent
This relation will be exactly satisfied if one define~[31]
$$
  f_a(x,{\mathbf k}_T^2,\mu^2)\vert_{{\mathbf k}_T^2 < \mu_0^2} = a(x,\mu_0^2) T_a(\mu_0^2,\mu^2). \eqno(25)
$$

\subsection{The CCFM unintegrated gluon distribution} \indent 

The CCFM gluon density has been obtained~[38] 
from the numerical solution of the CCFM equation. 
The function $f_g(x,{\mathbf k}_T^2,\mu^2)$ is determined
by a convolution of the non-perturbative starting
distribution $f_g^{(0)}(x)$ and the CCFM evolution kernel
denoted by $\tilde {\cal A}(x,{\mathbf k}_T^2,\mu^2)$:
$$
  f_g(x,{\mathbf k}_T^2,\mu^2) = \int {d x'\over x'} f_g^{(0)}(x') \tilde {\cal A}\left({x\over x'},{\mathbf k}_T^2,\mu^2\right). \eqno(26)
$$

\noindent
In the perturbative evolution the gluon splitting function
$P_{gg}(z)$ including nonsingular terms
is applied, as it was described in~[39]. The input parameters in $f_g^{(0)}(x)$
were fitted to reproduce the proton structure functions $F_2(x,Q^2)$.
An acceptable fit to the measured $F_2$ values was obtained~[38] with
$\chi^2/ndf = 1.83$ using statistical and uncorrelated systematic
uncertainties (compare to $\chi^2/ndf \sim 1.5$ in the collinear approach
at NLO).

\section{Numerical results} \indent

We are now in a position to present our numerical results.
First we describe the theoretical uncertainties of
our consideration.

Except the unintegrated parton distributions in a 
proton $f_q(x,{\mathbf k}_T^2,\mu^2)$,
there are several parameters which determined the overall 
normalization factor of the calculated $W^\pm/Z^0$ cross sections: 
the quark masses $m_1$ and $m_2$ and the factorization and 
renormalization scales $\mu_F$ and $\mu_R$
(the first of them is related to the evolution of the parton distributions, 
the other is responsible for the strong coupling constant).
In the numerical calculations the masses of light 
quarks were set to be equal to $m_u = 4.5$~MeV, $m_d = 8.5$~MeV, 
$m_s = 155$~MeV and the 
charmed quark mass was set to $m_c = 1.5$~GeV. 
We have checked that uncertainties which come 
from these quantities are negligible in comparison to the uncertainties
connected with the scale and/or the unintegrated parton densities.
As it is often done, we choose the 
renormalization and factorization scales to be equal: 
$\mu_R = \mu_F = \mu = \xi m_T$ (transverse mass of the
produced vector boson).
In order to investigate the scale dependence of our 
results we vary the scale parameter
$\xi$ between $1/2$ and~2 about the default value $\xi = 1$.
For completeness, we set $m_W = 80.403$~GeV, $m_Z = 91.1876$~GeV,
$\sin^2 \theta_W = 0.23122$ and use the LO formula for the strong 
coupling constant $\alpha_s(\mu^2)$ with $n_f = 4$ 
active quark flavors at 
$\Lambda_{\rm QCD} = 200$~MeV (so that $\alpha_s(M_Z^2) = 0.1232$).
Note that we use a special choice $\Lambda_{\rm QCD} = 130$~MeV 
in the case of CCFM gluon ($\alpha_s(M_Z^2) = 0.1187$), 
as it was originally proposed in~[38].

Before we proceed to the numerical results,
we would like to comment on the effect of the
higher order QCD contributions~[30]. It is well-known
that the leading-order $k_T$-factorization approach
naturally includes a large part of them\footnote{See, for example,
review~[29] for more details.}.
It is a corrections which are kinematic in nature 
arising from the real parton emission during the
evolution cascade. Another part of high-order contributions comes from
the logarithmic loop corrections which have
already been included in the Sudakov form factors~(22) and~(23).
However, there are also the non-logarithmic loop corrections,
arising, for example, from the gluon vertex corrections to Fig.~2.
To take into account these contributions we will
use the approach proposed in~[30]. It was demonstrated
that main part of the non-logarithmic loop corrections can be
absorbed in the so-called K-factor given by the expression
$$
  K(q + \bar q^\prime \to W^\pm/Z^0) \simeq \exp \left[C_F {\alpha_s(\mu^2)\over 2 \pi} \pi^2 \right], \eqno(27)
$$

\noindent
where color factor $C_F = 4/3$. A particular choice
$\mu^2 = {\mathbf p}_T^{4/3} m^{2/3}$ has been proposed~[22, 30]
to eliminate sub-leading logarithmic terms.
We choose this scale to evaluate the strong coupling constant 
$\alpha_s(\mu^2)$ in~(27).

We begin the discussion by presenting a comparison
between the different contributions to the $W^\pm/Z^0$
total cross section. The solid, dashed and
dotted histograms in Figs.~3 --- 6 represent
the $g^* + g^* \to W^\pm/Z^0 + q \bar q^\prime$,
$q + q^* \to W^\pm/Z^0 + q^\prime$ and 
$q + \bar q^\prime \to W^\pm/Z^0$ contributions
to the rapidity distributions of gauge boson calculated
at the Tevatron (Figs.~3 and~4) and LHC conditions
(Figs.~5 and~6). It is important
that in the last two subprocesses we take into account only
the valence quarks within the usual collinear 
approximation. For illustration, we used here
the KMR unintegrated gluon density.
We found that the role of the gluon-gluon fusion
subprocess is greatly increased at the LHC energy:
it contributes only about 2 or 3\% of the valence quark 
component at the Tevatron and more than 40\% at 
the LHC. Moreover, in the last case it 
dominates over the valence contributions at the 
central rapidities. The contribution of the valence quark-antiquark 
annihilation subprocess is important at the Tevatron and 
gives only a few percents at the LHC energy.
As expected, the contribution of the $q + g^* \to W^\pm/Z^0 + q^\prime$
subprocess is significant in the forward rapidity region, $|y| > 2$.
At this point, we can conclude that the gluon-gluon
fusion becomes an important production mechanism 
at high energies and therefore should be taken into
account in the calculations. However, we would like to note that 
there is an additional contribution which is not included
in the simple decomposition scheme proposed above.
As it was mentioned above, in this scheme 
it was assumed that sea quarks appear only at last 
gluon splitting and there is no contribution from the 
quarks coming from the earlier steps of the evolution
(and we absorb last step of the 
gluon ladder into hard matrix element 
$g^* + g^* \to W^\pm/Z^0 + q \bar q^\prime$). 
It is not clear in advance, whether
the last gluon splitting dominates or not.
In order to model this additional component, we have
repeated the calculations using the KMR unintegrated quark 
densities~(20) and the quark-antiquark annihilation 
$q + \bar q^\prime \to W^\pm/Z^0$ matrix element.
But in these evaluations we omited the last term and keep only sea quark 
in first term of~(20). Thus, we switch
off the pure gluon component of the sea quark distributions
and remove the valence quarks from the evolution ladder.
In this way only the contributions to the $f_q(x,{\mathbf k}_T^2,\mu^2)$ 
originating from the earlier (involving quarks) evolution steps
are taken into account. 
So, the dash-dotted histograms in Figs.~3 ---~6 represent
the results of our calculations. We have found the
significant (by about of 50\%) enhancement of the cross sections
at both the Tevatron and LHC conditions.
Therefore in all calculations below we will consider this
mechanism as an additional production one.
Finally, taking into account all described above components,
we can conclude that the gluon-gluon fusion contributes about
$\sim 1$\% to the total cross section at Tevatron and up to $\sim 25$\%
at LHC energies. 

Now we turn to the transverse momentum distributions of
the $W^\pm$ and $Z^0$ bosons.
The experimental data for the transverse momentum
distributions come from both the D$\oslash$~[8, 9] and CDF~[4] 
collaborations at Tevatron. These data were obtained at center-of-mass
energy $\sqrt s = 1800$~GeV. Measurements were made for $W \to l\nu$
and $Z \to l^+l^-$ decays; so that we should multiply our
theoretical predictions by the relevant branching fractions 
$f(W \to l\nu)$ and $f(Z \to l^+l^-)$. 
These branching fractions 
were set to $f(W \to l\nu) = 0.1075$ and $f(Z \to l^+l^-) = 0.03366$~[40].
In Figs.~7 --- 9 we display a comparison of the calculated 
differential cross sections $d\sigma/dp_T$ of the
$W^\pm$ and $Z^0$ boson production
with the experimental data~[4, 8, 9] in the low $p_T$ region, namely $p_T < 20$~GeV.
Next, in Figs.~10 --- 12,
we demonstrate the $W^\pm$ and $Z^0$ transverse momentum
distributions in the intermediate and high $p_T$ regions.
Additionally, in Figs.~13 and 14, we plot the normalized 
differential cross section
$(1/\sigma)\,d\sigma/dp_T$ of the $W^\pm$ boson production.
The solid and dashed histograms correspond to the 
results obtained with the CCFM and KMR unintegrated gluon densities, 
respectively. All contributions discussed above are taken
into account. The dotted histograms were obtained using
the quark-antiquark annihilation matrix element 
convoluted with the KMR unintegrated quark distributions
in a proton (in this case the transverse mometum of the 
produced vector boson is defined by the transverse momenta
of the incoming quarks).
We found an increase in the cross section calculated
in the proposed decomposition scheme (where only the unintegrated 
gluon densities used).
In this scheme, we obtain that both the CCFM and KMR gluon distributions
reproduce well the Tevatron data within the uncertainties,
although the KMR gluon tends to slightly underestimate
the data in the low $p_T$ region. 
The difference between the solid and dashed histograms in Figs.~7 --- 14
is due to different behaviour of the CCFM and KMR gluon densities.
The predictions based of the 
quark-antiquark annihilation subprocess lie
below the experimental data but agree with them in shape.
This observation coincides with the one from~[30] where
an additional factor of about 1.2 was introduced
to eliminate the visible disagreement between the data and 
theory\footnote{In Ref.~[30] authors have explained the origin of 
this extra factor by the fact that the input
parton densities (used to determine the unintegrated ones)
should themselves be determined from data using
the appropriate non-collinear formalism.}. 

An additional possibility to distinguish
the two calculation schemes comes from the
studying of the ratio of the $W^\pm$ and $Z^0$ boson 
cross sections. In fact, since $W^\pm$ and $Z^0$ production
properties are very similar, as
the transverse momentum of the vector boson becomes
smaller, the radiative corrections affecting the individual distributions and 
the cross sections of hard process are
factorized and canceled in this ratio.
Therefore the results of calculation of this ratio 
in the decomposition scheme (where
the ${\cal O}(\alpha \alpha_s)$ and ${\cal O}(\alpha \alpha_s^2)$ 
subprocesses are taken into account) and the predictions 
based on the ${\cal O}(\alpha)$
quark-antiquark annihilation should
differ from each other at moderate and high $p_T$ values.
This fact is clearly illustrated in Fig.~15 where the ratio of
$W^\pm$ and $Z^0$ cross sections as a function of the
transverse momentum is displayed.
As it was expected, there is practically no difference
between all plotted histograms in the low $p_T$ region.

As a final point of our study, we discuss the scale dependence of
our results. In Figs.~16 and~17 we show the total
cross section of the $W^\pm$ and $Z^0$ boson 
production as a function of the total center-of-mass 
energy $\sqrt s$. Here, the solid and dotted histograms 
correspond to the results obtained with the CCFM 
and KMR unintegrated gluon densities, respectively. The
the upper and lower dashed histograms 
correspond to the scale variations in the CCFM gluon density 
as it was described above. We find that the
scale uncertainties are the same order approximately 
as the uncertainties coming from the unintegrated
gluon distributions. This fact is the typical one for the
leading-order $k_T$-factorization calculations. 
Our predictions for the $W^\pm$ and $Z^0$ boson 
total cross section agree well with the data
in a wide $\sqrt s$ range.

\section{Conclusions} \indent 

We have studied the production of 
electroweak gauge bosons in hadronic collisions at high energies
in the $k_T$-factorization approach of QCD.
 Our consideration is based on the scheme which provides solid 
theoretical grounds 
for adequately taking into account the effects of initial parton 
momentum. The central part of our derivation is the off-shell gluon-gluon 
fusion subprocess $g^* + g^* \to W^\pm/Z^0 + q \bar q^\prime$. 
At the price of considering the corresponding matrix element 
rather than $q + \bar q^\prime \to W^\pm/Z^0$ one, 
we have reduced the problem of unknown unintegrated quark distributions to the 
problem of gluon distributions. 
This way enables us with making comparisons between the different parton 
evolution schemes and parametrizations of parton densities. 
Since the gluons are only responsible for the appearance of sea, but 
not valence quarks, the contribution from the valence quarks has been 
calculated separately. Having in mind that the valence quarks are only 
important at large $x$, where the traditional DGLAP evolution is accurate and 
reliable, we have calculated this contribution within the usual collinear 
scheme based on $q + g^* \to W^\pm/Z^0 + q^\prime$ and 
$q + \bar q^\prime \to W^\pm/Z^0$ partonic subprocesses 
and on-shell parton densities. 

We have studied in detail the different production 
mechanisms of $W^\pm$ and $Z^0$ bosons. 
We find that the off-shell gluon-gluon fusion
gives $\sim 1$\% and $\sim 25$\% contributions
to the inclusive $W^\pm/Z^0$ production cross sections
at the Tevatron and LHC.
Specially we simulate the contribution from
the quarks involved into the earlier steps of the 
evolution cascade (i.e., into the second-to-last, third-to-last and 
other gluon splittings) and find that these quarks
play an important role at both the Tevatron and LHC energies. 
It was demonstrated that corresponding corrections
should be taken into accout in the numerical calculations 
within the $k_T$-factorization approach.

We have calculated the total and differential $W^\pm$ and $Z^0$ 
production cross sections and have made comparisons with the Tevatron 
data. In the numerical analysis we have used 
the unintegrated gluon densities obtained from the  
CCFM evolution equation and from the KMR prescription.
Our numerical results agree well with
the experimental data.

When the present paper was ready for publication,
we have learned about the results obtained by
M.~Deak and F.~Schwennsen~[41], who used the same theoretical approach,
but focused attention on slightly different aspects of the problem.
These authors concentrate on the associated $W^\pm/Z^0$ production
with heavy quark pairs (mainly on the $Z b\bar b$ final 
state), where the gluon-gluon fusion subprocess dominates.
In additional to that, we consider quark subprocesses, which 
are important for inclusive $W^\pm/Z^0$ production. We show that the 
experimental data can be described with taking quark contributions into account.

\section{Acknowledgements} \indent 

We thank H.~Jung for his encouraging interest, very helpful discussions
and for providing the CCFM code for 
unintegrated gluon distributions. We
thank M.~Deak and F.~Schwennsen for letting us
know about the results of their work before
publication and useful remarks.
The authors are very grateful to 
DESY Directorate for the support in the 
framework of Moscow --- DESY project on Monte-Carlo
implementation for HERA --- LHC.
A.V.L. was supported in part by the grants of the president of 
Russian Federation (MK-438.2008.2) and Helmholtz --- Russia
Joint Research Group.
Also this research was supported by the 
FASI of Russian Federation (grant NS-8122.2006.2)
and the RFBR fundation (grant 08-02-00896-a).

\section{Appendix A} \indent 

Here we present the compact analytic expressions for the 
cross section of the $W^\pm/Z^0$ production
via $q + \bar q^\prime \to W^\pm/Z^0$ subprocess
in the $k_T$-factorization approach. 
Let us define the transverse momenta and azimuthal
angles of incoming quark $q$ and antiquark $\bar q^\prime$ as 
${\mathbf k}_{1T}$ and ${\mathbf k}_{2T}$ and
$\phi_1$ and $\phi_2$, respectively. 
The produced vector boson has the transverse momentum
${\mathbf p}_{T}$ (${\mathbf p}_{T} = {\mathbf k}_{1T} + {\mathbf k}_{2T}$) 
and center-of-mass rapidity $y$.
The $W^\pm/Z^0$ production cross section can be written as
$$
  \displaystyle \sigma(p + \bar p \to W^\pm/Z^0 + X) = \sum_{q} \int {2 \pi \over (x_1 x_2 s)^2} |\bar {\cal M}(q + \bar q^\prime \to W^\pm/Z^0)|^2 \times \atop 
  \displaystyle \times f_q(x_1,{\mathbf k}_{1 T}^2,\mu^2) f_q(x_2,{\mathbf k}_{2 T}^2,\mu^2) d{\mathbf k}_{1 T}^2 d{\mathbf k}_{2 T}^2 dy {d\phi_1\over 2\pi} {d\phi_2\over 2\pi}, \eqno(A.1)
$$

\noindent 
where $f_q(x,{\mathbf k}_{T}^2,\mu^2)$ is the unintegrated quark
distributions given by~(20). In the high-energy limit the 
fractions $x_1$ and $x_2$ of the
initial proton's longitudinal momenta are given by
$$
  x_1 \sqrt s = m_T\,e^y, \quad x_2 \sqrt s = m_T \, e^{-y}. \eqno(A.2)
$$

\noindent 
where $m_T$ is the transverse mass of the vector boson.
The squared matrix element $|\bar {\cal M}(q + \bar q^\prime \to W^\pm)|^2$ 
summed over final polarization states and averaged over initial ones is
$$
  |\bar {\cal M}(q + \bar q^\prime \to W^\pm)|^2 = - {e^2\over 72 m_W^2 \sin^2 \theta_W} \left[(m_1^2 - m_2^2)^2 + m_W^2 (m_1^2 + m_2^2) - 2 m_W^4\right], \eqno(A.3)
$$

\noindent 
where $m_1$ and $m_2$ are the masses of incoming quarks.
In the case of $Z^0$ boson production,
the squared matrix element $|\bar {\cal M}(q + \bar q \to Z^0)|^2$ 
summed over final polarization states and averaged over initial ones is
$$
  \displaystyle |\bar {\cal M}(q + \bar q \to Z^0)|^2 = {2 e^2\over 9 \sin^2 2 \theta_W} \times \atop
  \displaystyle \times \left[ (m_Z^2 - m^2)\left[I_3^{(q)}\right]^2 + 2 e_q (2 m^2 + m_Z^2) \sin^2 \theta_W \left(e_q \sin^2\theta_W - I_3^{(q)}\right) \right], \eqno(A.4)
$$

\noindent 
where $m$, $e_q$ and $I_3^{(q)}$ is the mass, fractional electric charge and
weak isospin of incoming quark. Note that there is no obvious 
dependence on the transverse momenta of the initial quark and antiquark.
However, this dependence is present because 
the true off-shell kinematics is used. 
In particular, the incident parton momentum fractions
$x_1$ and $x_2$ have some ${\mathbf k}_{T}$ 
dependence. If we take the limit ${\mathbf k}_{1 T}^2 \to 0$
and ${\mathbf k}_{2 T}^2 \to 0$,
then we recover the relevant expression in the standard 
collinear approximation of QCD.

\section{Appendix B} \indent 

Here we present the analytic expressions for the 
cross section of the $W^\pm/Z^0$ production
via $q + g^* \to W^\pm/Z^0 + q^\prime$ subprocess
in the $k_T$-factorization approach. 
Let us define the transverse momenta and azimuthal
angles of incoming quark and off-shell gluon as 
${\mathbf k}_{1T}$ and ${\mathbf k}_{2T}$ and
$\phi_1$ and $\phi_2$, respectively. 
In the following, $\hat s$, $\hat t$ and $\hat u$ are usual Mandelstam 
variables for $2 \to 2$ subprocess.
The $W^\pm/Z^0$ production cross section can be written as follows:
$$
  \displaystyle \sigma(p + \bar p \to W^\pm/Z^0 + X) = \sum_{q} \int {1\over 16\pi (x_1 x_2 s)^2} |\bar {\cal M}(q + g^* \to W^\pm/Z^0 + q^\prime)|^2 \times \atop 
  \displaystyle \times f_q(x_1,{\mathbf k}_{1 T}^2,\mu^2) f_g(x_2,{\mathbf k}_{2 T}^2,\mu^2) d{\mathbf k}_{1 T}^2 d{\mathbf k}_{2 T}^2 d{\mathbf p}_{T}^2 dy dy^\prime {d\phi_1\over 2\pi} {d\phi_2\over 2\pi}, \eqno(B.1)
$$

\noindent
where $y^\prime$ is the rapidity of the final quark $q^\prime$. The
fractions $x_1$ and $x_2$ of the
initial proton's longitudinal momenta are given by
$$
  x_1 \sqrt s = m_{T}\,e^y + m_{T}^\prime \,e^{y^\prime}, \quad x_2 \sqrt s = m_{T}\, e^{-y} + m_{T}^\prime \,e^{-y^\prime}. \eqno(B.2)
$$

\noindent 
where $m_T$ and $m_T^\prime$ are the transverse masses of the 
vector boson and final quark $q^\prime$.
If we take the limit ${\mathbf k}_{1 T}^2 \to 0$ and ${\mathbf k}_{2 T}^2 \to 0$,
then we recover the relevant expression in the usual 
collinear approximation.
The squared matrix elements $|\bar {\cal M}(q + g^* \to W^\pm + q^\prime)|^2$ 
and $|\bar {\cal M}(q + g^* \to Z^0 + q)|^2$ summed 
over final polarization states and averaged over initial ones are
$$
  |\bar {\cal M}(q + g^* \to W^\pm + q^\prime)|^2 = {e^2 g^2 \over 192 \sin^2 \theta_W} {F_W \over (m_1^2 - \hat s)^2 (m_2^2 - \hat t)^2 m_W^2}, \eqno(B.3)
$$
$$
  |\bar {\cal M}(q + g^* \to Z^0 + q)|^2 = {2 e^2 g^2 \over 3 \sin^2 2 \theta_W} {F_Z \over (m^2 - \hat s)^2 (m^2 - \hat t)^2 m_Z^2}, \eqno(B.4)
$$

\noindent where
$$
  F_W = -8 (m_1^8 (3 m_2^2 - \hat t) + m_1^6 (m_2^4 + m_2^2 (2 m_W^2 - 5 \hat s - 7 \hat t) + \hat t (\hat s + 2 \hat t)) +
$$
$$ 
  m_1^4 (m_2^6 + m_2^4 (8 m_W^2 - 3 (\hat s + \hat t)) + m_2^2 (-6 m_W^4 + 3 \hat s^2 + 13 \hat s \hat t + 5 \hat t^2 - 6 m_W^2 (\hat s + \hat t)) - 
$$
$$
  \hat t (-2 m_W^4 - 2 m_W^2 \hat s + (\hat s + \hat t)^2)) + m_1^2 (3 m_2^8 + m_2^6 (2 m_W^2 - 7 \hat s - 5 \hat t) + 
$$
$$
  m_2^4 (-6 m_W^4 + 5 \hat s^2 + 13 \hat s \hat t + 3 \hat t^2 - 6 m_W^2 (\hat s + \hat t)) + m_2^2 (4 m_W^6 - \hat s^3 - 11 \hat s^2 \hat t - 11 \hat s \hat t^2 - \hat t^3 + 
$$
$$
  6 m_W^4 (\hat s + \hat t) + 6 m_W^2 (\hat s^2 + \hat t^2)) + \hat t (-4 m_W^6 + 2 m_W^4 \hat s + \hat s (\hat s + \hat t)^2 - 
$$
$$
  2 m_W^2 (2 \hat s^2 - \hat s \hat t + \hat t^2))) + \hat s (-m_2^8 + m_2^6 (2 \hat s + \hat t) + 2 m_W^2 \hat t (2 m_W^4 + \hat s^2 + \hat t^2 - 2 m_W^2 (\hat s + \hat t)) + 
$$
$$
  m_2^4 (2 m_W^4 + 2 m_W^2 \hat t - (\hat s + \hat t)^2) + m_2^2 (-4 m_W^6 + 2 m_W^4 \hat t + \hat t (\hat s + \hat t)^2 - 
$$
$$
  2 m_W^2 (\hat s^2 - \hat s \hat t + 2 \hat t^2))) + (m_1^8 + m_2^8 + m_1^6 (m_W^2 - 2 (\hat s + \hat t)) + m_2^6 (m_W^2 - 2 (\hat s + \hat t)) + 
$$
$$
  m_2^4 (-2 m_W^4 - 2 m_W^2 \hat t + (\hat s + \hat t)^2) + m_2^2 m_W^2 (5 \hat s^2 + 4 \hat s \hat t + \hat t^2 + m_W^2 (-8 \hat s + 4 \hat t)) - 
$$
$$
  2 m_W^2 (2 \hat s \hat t (\hat s + \hat t) + m_W^2 (\hat s^2 - 4 \hat s \hat t + \hat t^2)) + m_1^4 (-2 m_2^4 - 2 m_W^4 - 2 m_W^2 \hat s + (\hat s + \hat t)^2 + 
$$
$$
  m_2^2 (m_W^2 + 2 (\hat s + \hat t))) + m_1^2 (m_W^2 (\hat s^2 + 4 m_W^2 (\hat s - 2 \hat t) + 4 \hat s \hat t + 5 \hat t^2) + m_2^4 (m_W^2 + 2 (\hat s + \hat t)) + 
$$
$$
  m_2^2 (8 m_W^4 - 6 m_W^2 (\hat s + \hat t) - 2 (\hat s + \hat t)^2))) ( - {\mathbf k}_{2T}^2 ) + 4 m_W^2 (m_1^2 - \hat s) (m_2^2 - \hat t) {\mathbf k}_{2T}^4), \eqno(B.5)
$$
$$
  F_Z = -2 e_q I_3^{(q)} m_Z^2 \sin^2 \theta_W (6 m^8 - \hat s \hat t (2 m_Z^4 + \hat s^2 + \hat t^2 - 2 m_Z^2 (\hat s + \hat t)) - 
$$
$$
  m^4 (2 m_Z^4 + 3 \hat s^2 + 14 \hat s \hat t + 3 \hat t^2 - 2 m_Z^2 (\hat s + \hat t)) + m^2 (\hat s^3 - 8 m_Z^2 \hat s \hat t + 7 \hat s^2 \hat t + 
$$
$$
  7 \hat s \hat t^2 + \hat t^3 + 2 m_Z^4 (\hat s + \hat t))) + 2 e_q^2 m_Z^2 \sin^4 \theta_W (6 m^8 - \hat s \hat t (2 m_Z^4 + \hat s^2 + \hat t^2 - 2 m_Z^2 (\hat s + \hat t)) - 
$$
$$
  m^4 (2 m_Z^4 + 3 \hat s^2 + 14 \hat s \hat t + 3 \hat t^2 - 2 m_Z^2 (\hat s + \hat t)) + m^2 (\hat s^3 - 8 m_Z^2 \hat s \hat t + 7 \hat s^2 \hat t + 7 \hat s \hat t^2 + \hat t^3 + 
$$
$$
  2 m_Z^4 (\hat s + \hat t))) + \left[I_3^{(q)}\right]^2 (-4 m^{10} + m^8 (-6 m_Z^2 + 8 (\hat s + \hat t)) - m_Z^2 \hat s \hat t (2 m_Z^4 + \hat s^2 + \hat t^2 - 
$$
$$
  2 m_Z^2 (\hat s + \hat t)) + m^6 (6 m_Z^4 - 5 \hat s^2 - 14 \hat s \hat t - 5 \hat t^2 + 6 m_Z^2 (\hat s + \hat t)) + 
$$
$$
  m^4 (-2 m_Z^6 + \hat s^3 + 7 \hat s^2 \hat t + 7 \hat s \hat t^2 + \hat t^3 - 4 m_Z^4 (\hat s + \hat t) - m_Z^2 (3 \hat s^2 + 2 \hat s \hat t + 3 \hat t^2)) + 
$$
$$
  m^2 (-2 m_Z^4 \hat s \hat t + 2 m_Z^6 (\hat s + \hat t) - \hat s \hat t (\hat s + \hat t)^2 + m_Z^2 (\hat s^3 + \hat s^2 \hat t + \hat s \hat t^2 + \hat t^3))) +
$$
$$ 
  m_Z^2 (2 e_q I_3^{(q)} \sin^2 \theta_W (2 m^4 (m_Z^2 - \hat s - \hat t) - 2 \hat s \hat t (\hat s + \hat t) - m_Z^2 (\hat s^2 - 4 \hat s \hat t + \hat t^2) - 
$$
$$
  2 m^2 (-4 \hat s \hat t + m_Z^2 (\hat s + \hat t))) + 2 e_q^2 \sin^4 \theta_W (2 \hat s \hat t (\hat s + \hat t) + 2 m^4 (-m_Z^2 + \hat s + \hat t) + 
$$
$$
  m_Z^2 (\hat s^2 - 4 \hat s \hat t + \hat t^2) + 2 m^2 (-4 \hat s \hat t + m_Z^2 (\hat s + \hat t))) + \left[I_3^{(q)}\right]^2 (-2 m^6 + 2 \hat s \hat t (\hat s + \hat t) + 
$$
$$
m_Z^2 (\hat s^2 - 4 \hat s \hat t + \hat t^2) + m^4 (-2 m_Z^2 + 4 (\hat s + \hat t)) + m^2 (-3 \hat s^2 - 4 \hat s \hat t - 3 \hat t^2 + 
$$
$$
  2 m_Z^2 (\hat s + \hat t)))) ( - {\mathbf k}_{2T}^2 ) - 2 m_Z^2 (m^2 - \hat s) (\left[I_3^{(q)}\right]^2 - 
$$
$$
  2 e_q I_3^{(q)} \sin^2 \theta_W + 2 e_q^2 \sin^4 \theta_W) (m^2 - \hat t) {\mathbf k}_{2T}^4. \eqno(B.6)
$$

\newpage

\begin{figure}
\begin{center}
\epsfig{figure=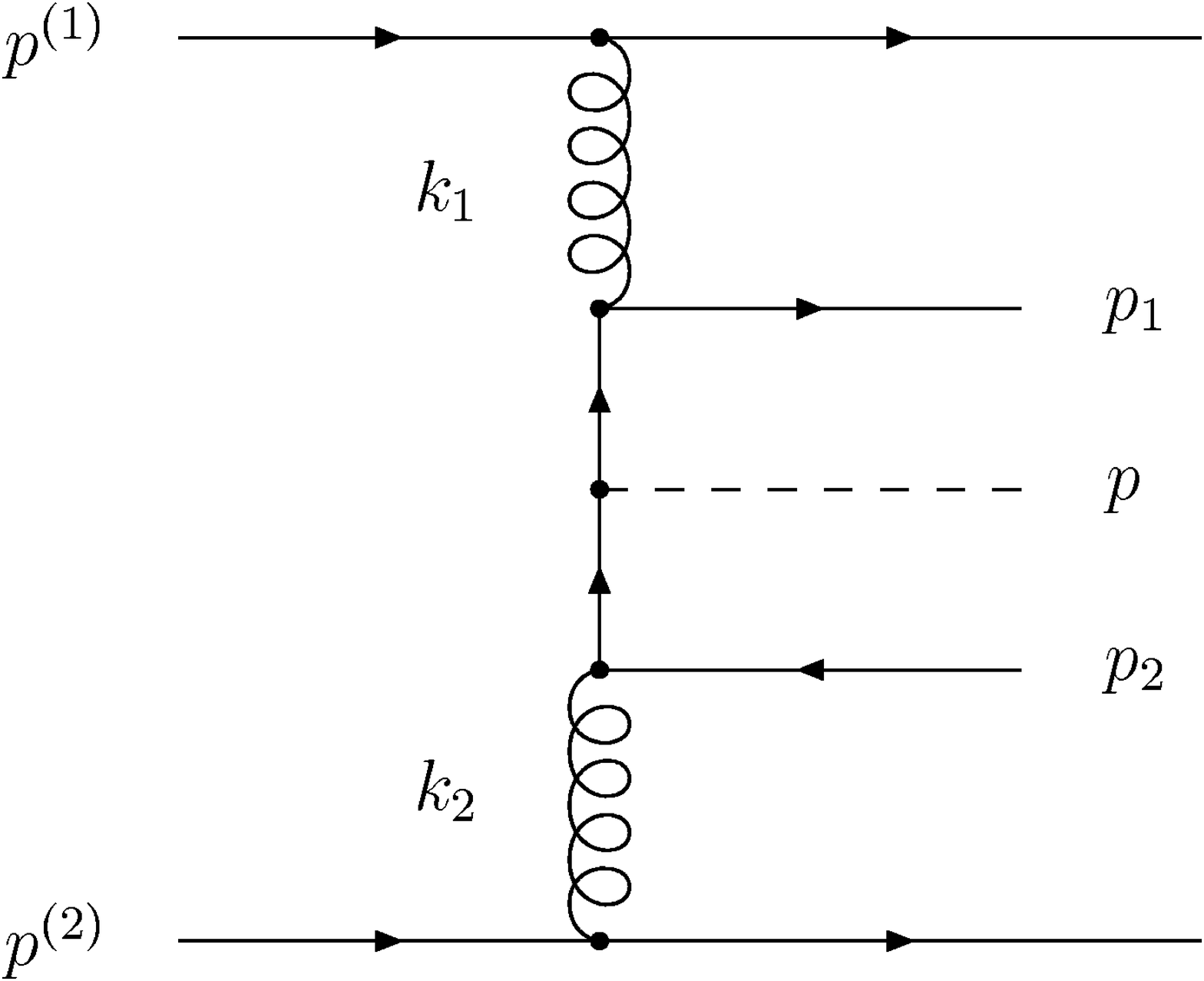, width = 8cm}
\caption{Kinematics of the $g^* + g^* \to W^\pm/Z^0 + q q^\prime$ process.}
\label{fig1}
\end{center}
\end{figure}

\newpage

\begin{figure}
\begin{center}
\epsfig{figure=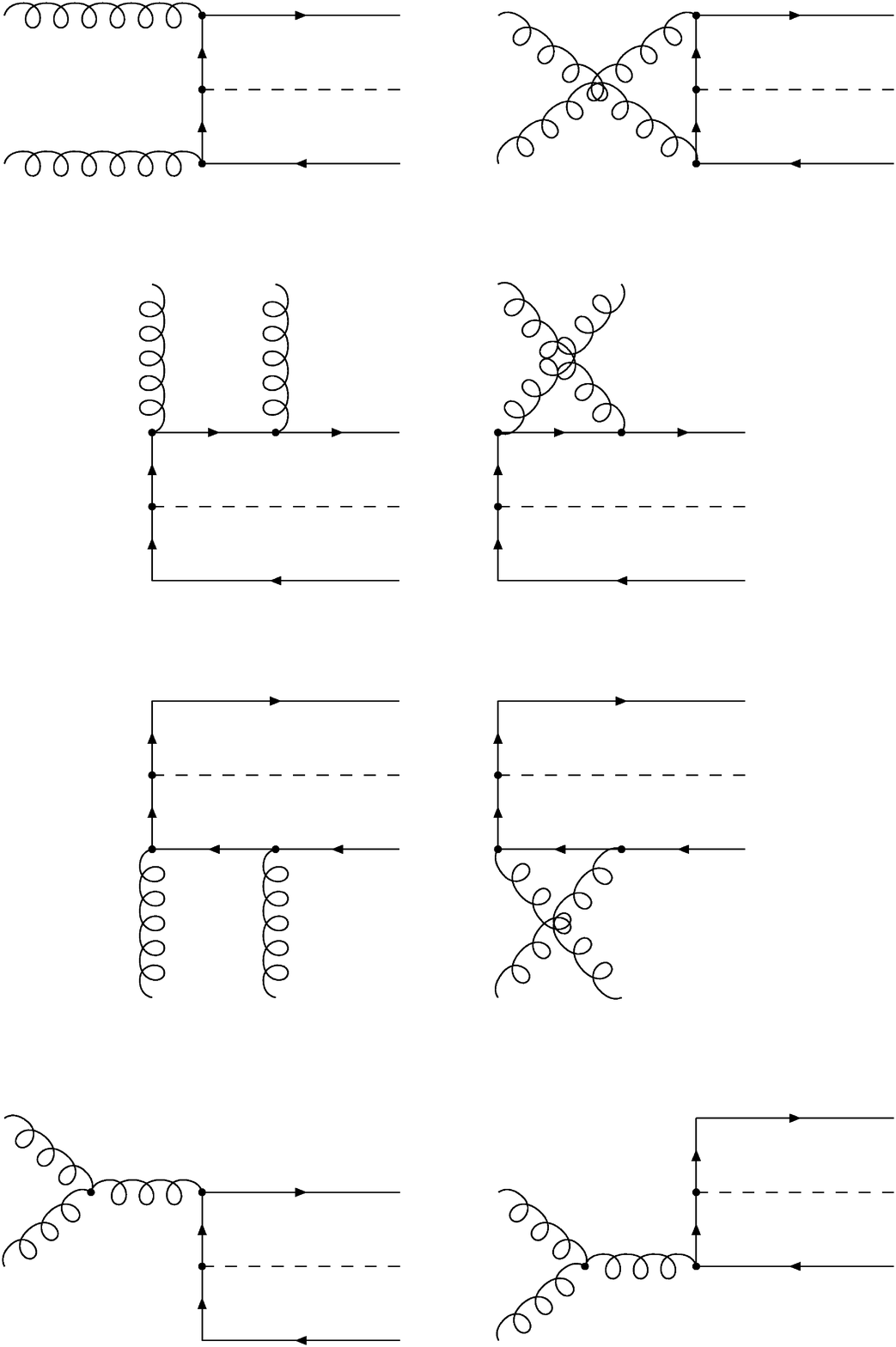, width = 14cm}
\caption{Feynman diagrams which describe the partonic
subprocess $g^* + g^* \to W^\pm/Z^0 + q q^\prime$ at the leading order in $\alpha_s$ and $\alpha$.}
\label{fig2}
\end{center}
\end{figure}

\newpage

\begin{figure}
\begin{center}
\epsfig{figure=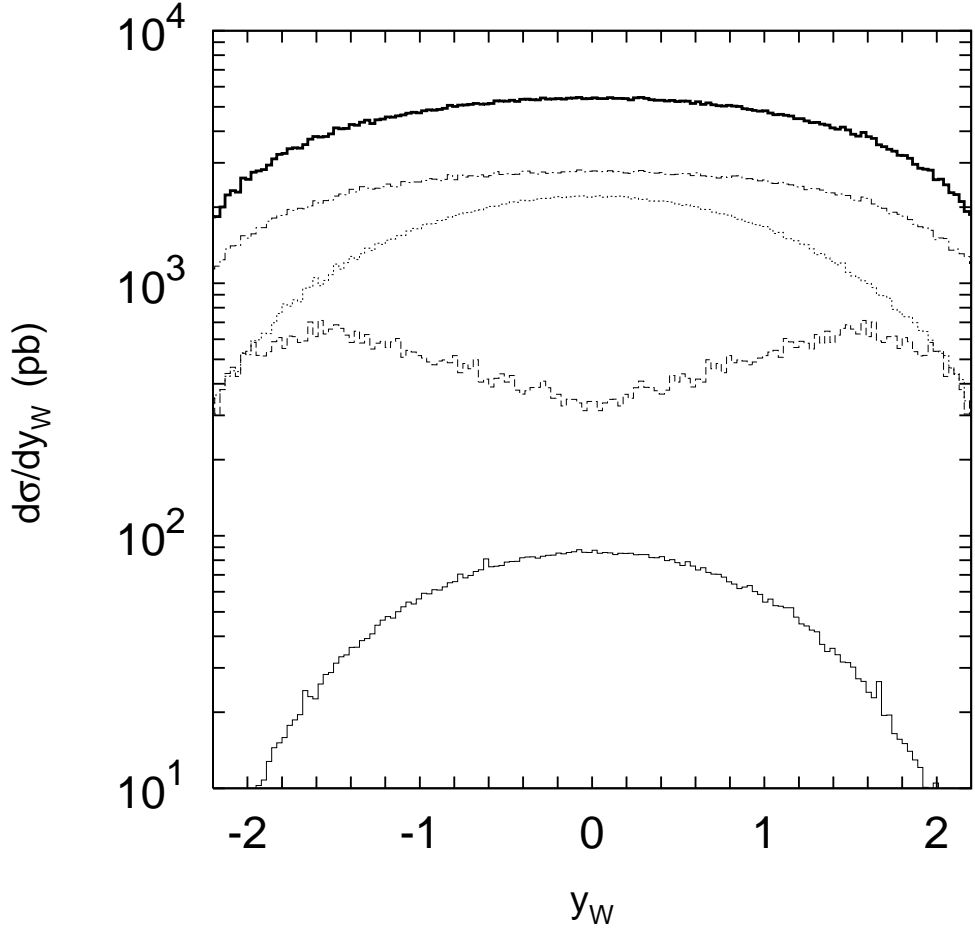, width = 18cm}
\caption{Different contributions to the
inclusive $W^\pm$ boson production at the Tevatron conditions.
The solid, dashed and dotted histograms correspond to the 
$g^* + g^* \to W^\pm + q \bar q^\prime$,
$q + g^* \to W^\pm + q^\prime$ and 
$q + \bar q^\prime \to W^\pm$ subprocesses.
In the last two cases only the valence quarks
are taken into account. The dash-dotted histogram
represents contribution from the quarks coming 
from the earlier steps of the evolution.
The thick solid histogram represents the sum of all contributions.
The KMR unintegrated parton densities in a proton are used.}
\end{center}
\label{fig3}
\end{figure}

\newpage

\begin{figure}
\begin{center}
\epsfig{figure=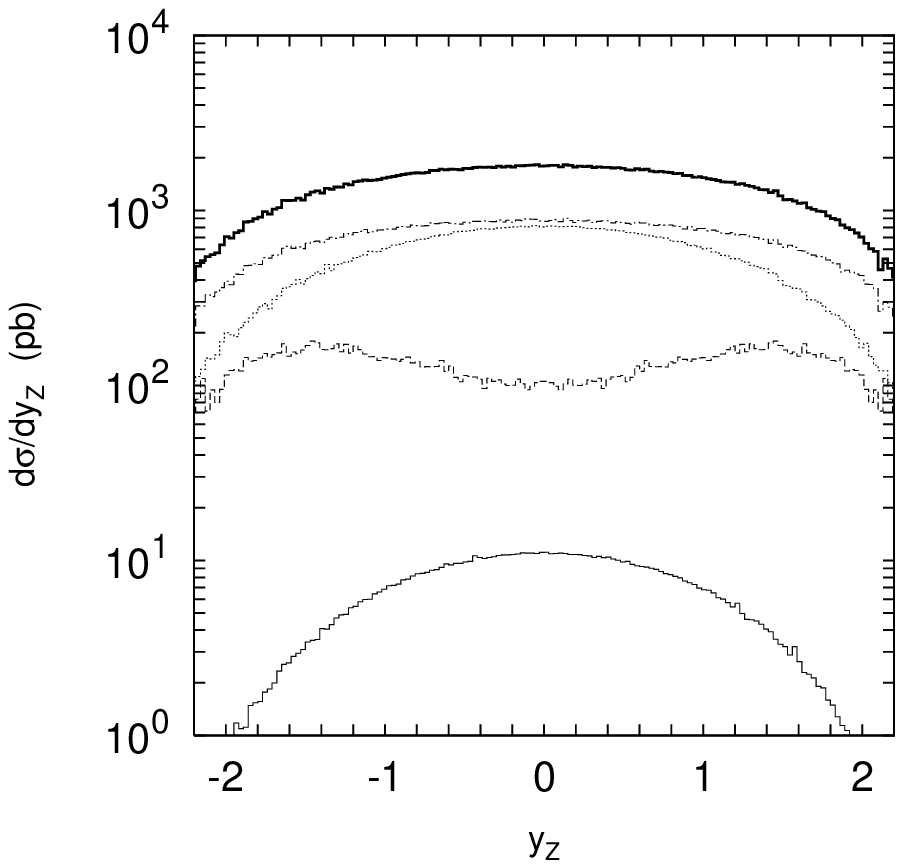, width = 18cm}
\caption{Different contributions to the
inclusive $Z^0$ boson production at the Tevatron conditions.
Notations of histograms are the same as in Fig.~3.}
\end{center}
\label{fig4}
\end{figure}

\newpage

\begin{figure}
\begin{center}
\epsfig{figure=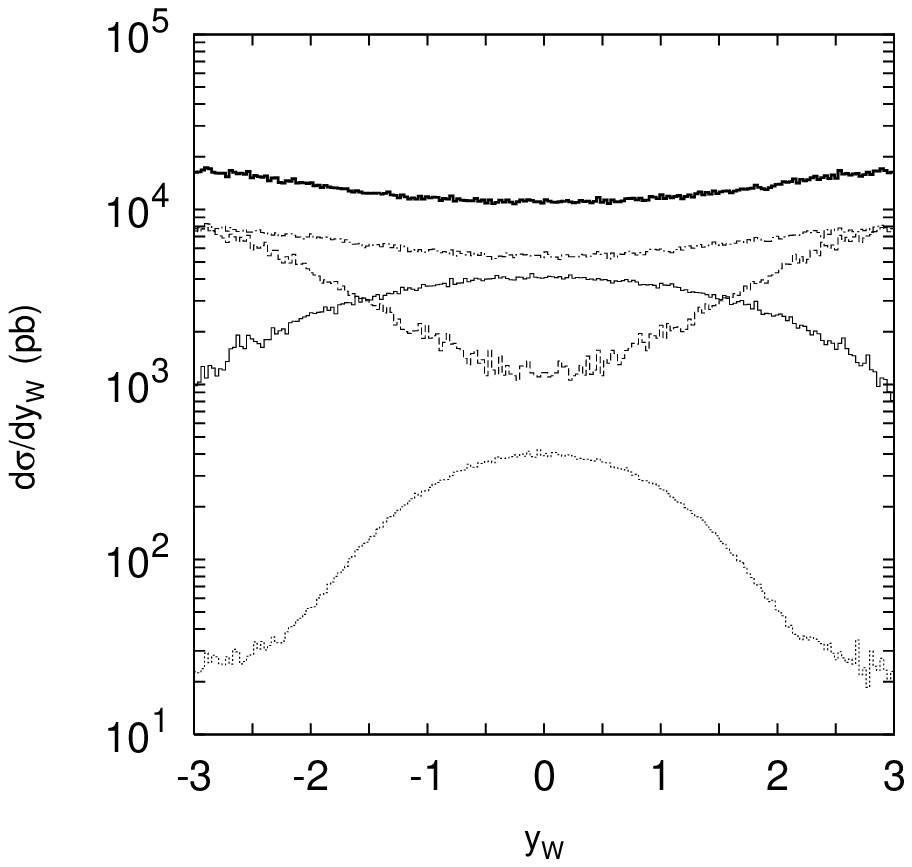, width = 18cm}
\caption{Different contributions to the
inclusive $W^\pm$ boson production at the LHC conditions.
Notations of histograms are the same as in Fig.~3.}
\end{center}
\label{fig5}
\end{figure}

\newpage

\begin{figure}
\begin{center}
\epsfig{figure=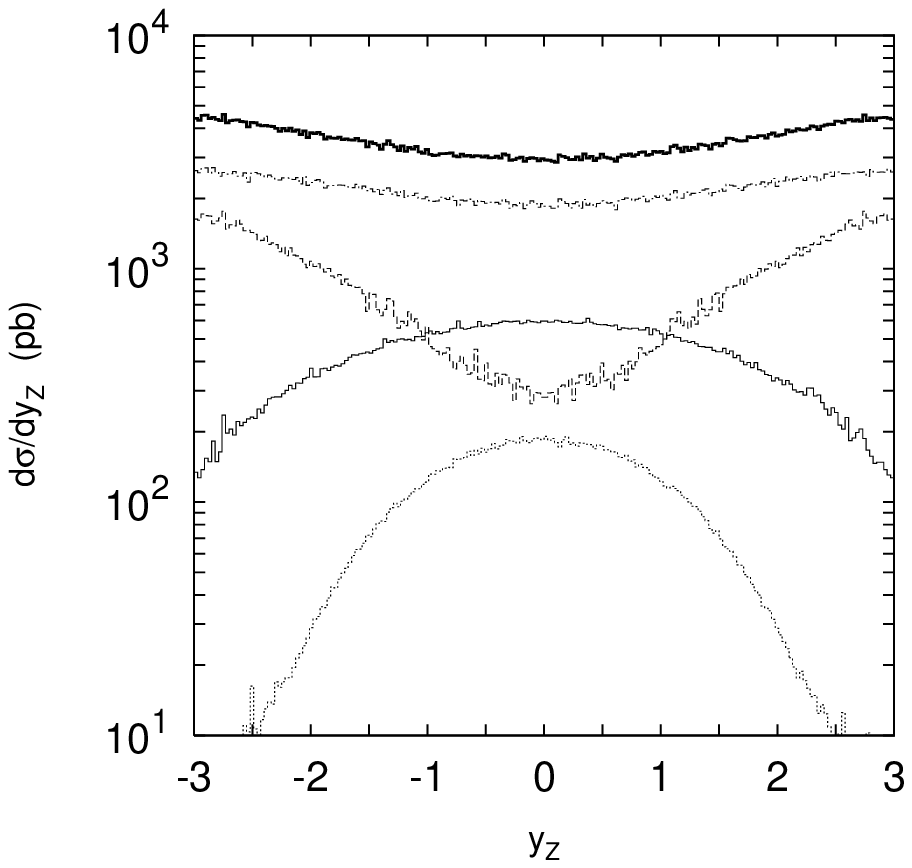, width = 18cm}
\caption{Different contributions to the
inclusive $Z^0$ boson production at the LHC conditions.
Notations of histograms are the same as in Fig.~3.}
\end{center}
\label{fig6}
\end{figure}

\newpage

\begin{figure}
\begin{center}
\epsfig{figure=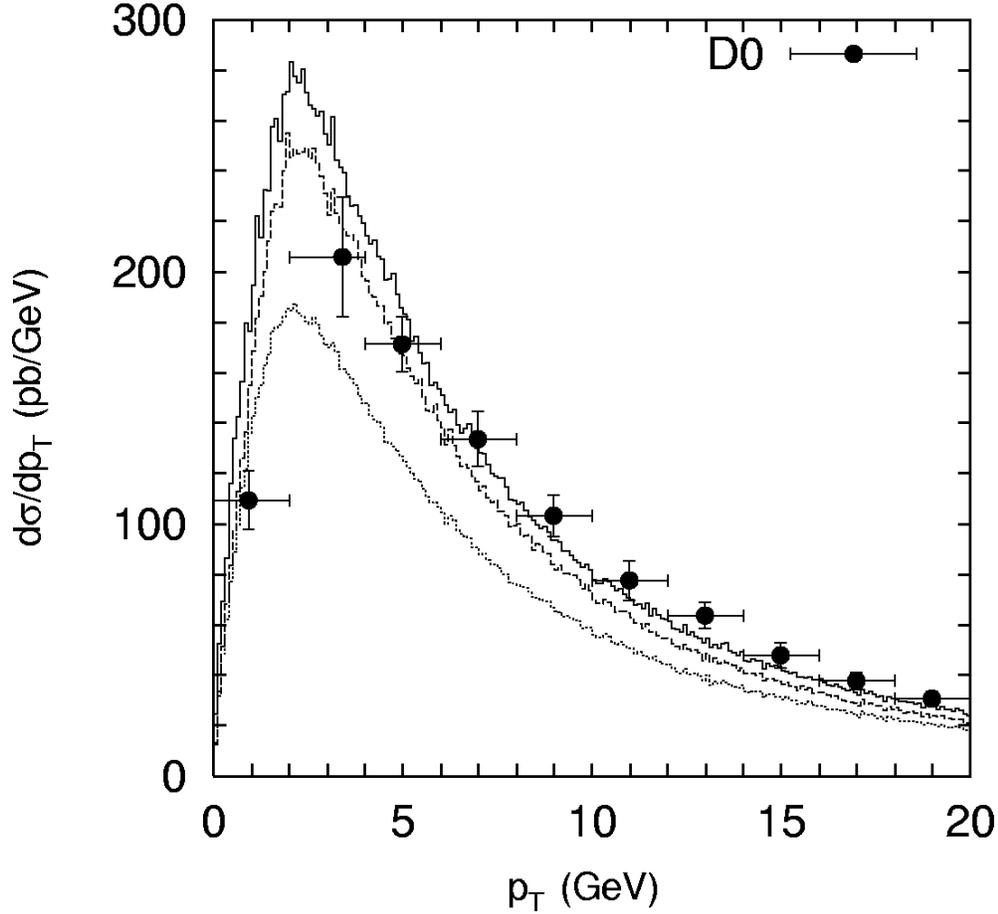, width = 18cm}
\caption{Transverse mometum distribution of
the $W^\pm$ boson production. The solid and dashed histograms correspond to the 
results obtained with the CCFM and KMR unintegrated gluon densities in 
a proton, respectively. The dotted histograms were obtained by using
the quark-antiquark annihilation matrix element 
convoluted with the KMR unintegrated quark distributions.
The cross sections times branching fraction $f(W \to l\nu)$ are shown.
The experimental data are from D$\oslash$~[9].}
\end{center}
\label{fig7}
\end{figure}

\newpage

\begin{figure}
\begin{center}
\epsfig{figure=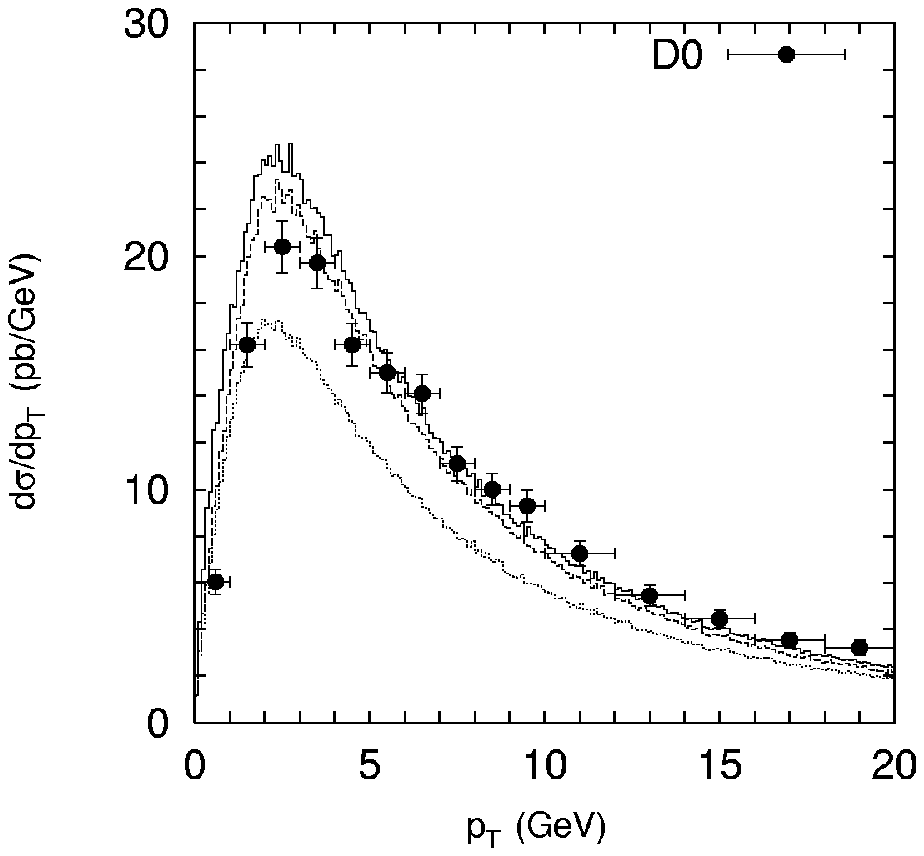, width = 18cm}
\caption{Transverse mometum distribution of
the $Z^0$ boson production. 
Notations of histograms are the same as in Fig.~7.
The cross sections times branching fraction $f(Z \to l^+l^-)$ are shown.
The experimental data are from D$\oslash$~[8].}
\end{center}
\label{fig8}
\end{figure}

\newpage

\begin{figure}
\begin{center}
\epsfig{figure=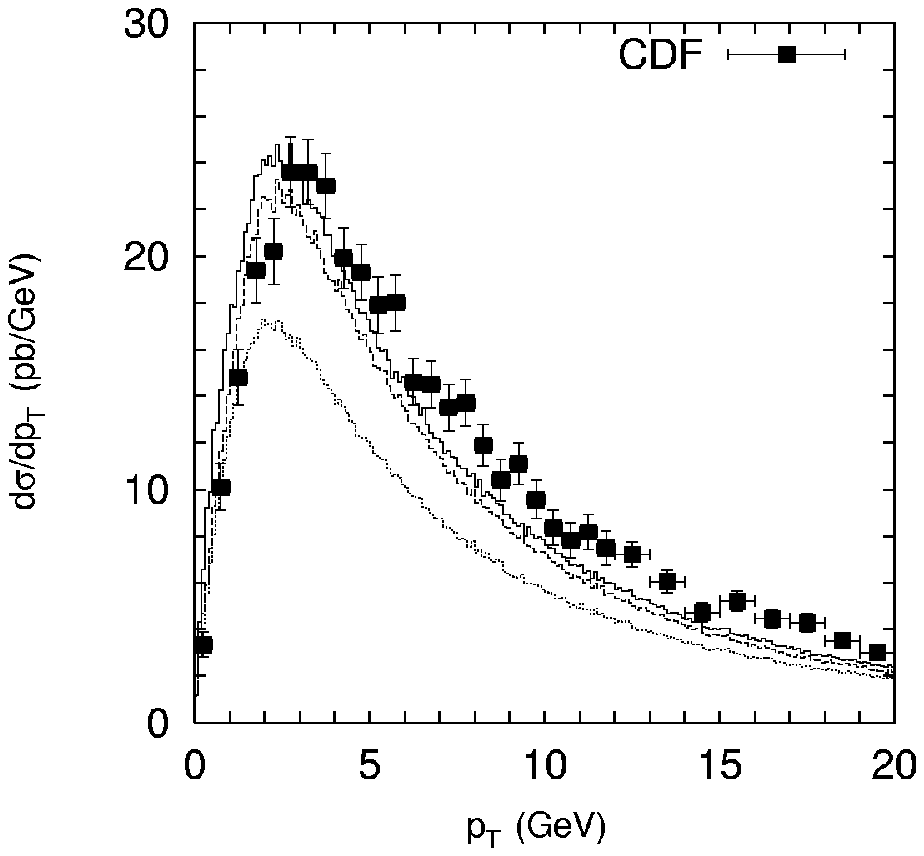, width = 18cm}
\caption{Transverse mometum distribution of
the $Z^0$ boson production. Notations of histograms are 
the same as in Fig.~7. 
The cross sections times branching fraction $f(Z \to l^+l^-)$ are shown.
The experimental data are from CDF~[4].}
\end{center}
\label{fig9}
\end{figure}

\newpage

\begin{figure}
\begin{center}
\epsfig{figure=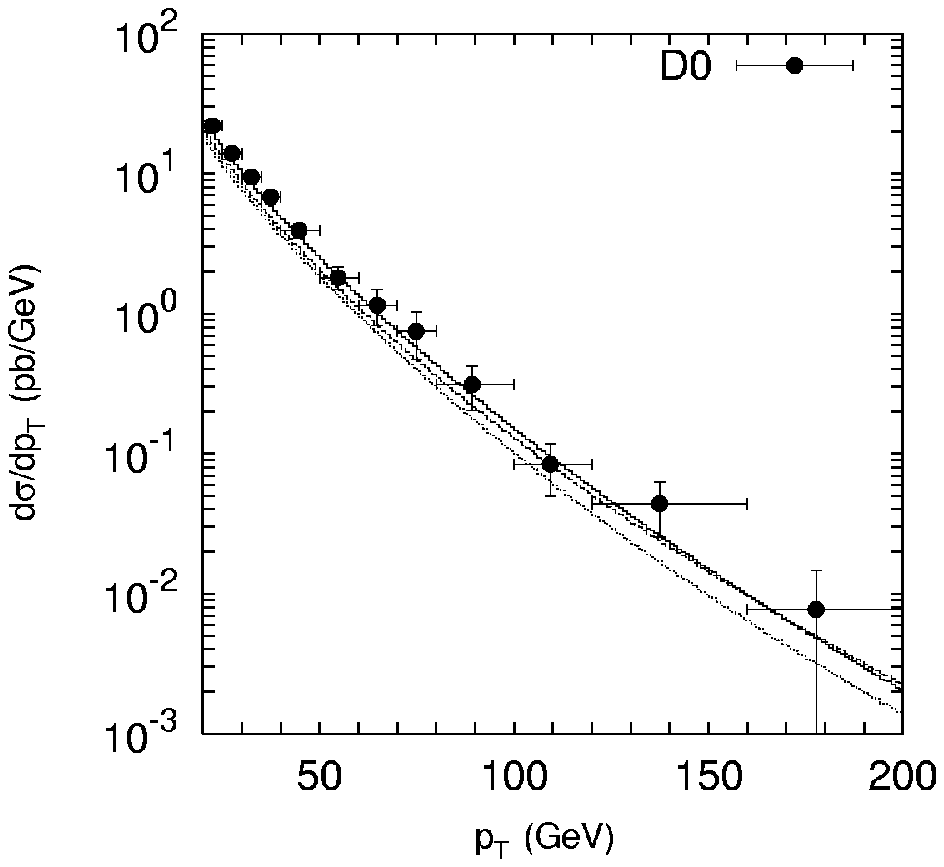, width = 18cm}
\caption{Transverse mometum distribution of
the $W^\pm$ boson production. Notations of histograms are 
the same as in Fig.~7. 
The cross sections times branching fraction $f(W \to l\nu)$ are shown.
The experimental data are from D$\oslash$~[9].}
\end{center}
\label{fig10}
\end{figure}

\newpage

\begin{figure}
\begin{center}
\epsfig{figure=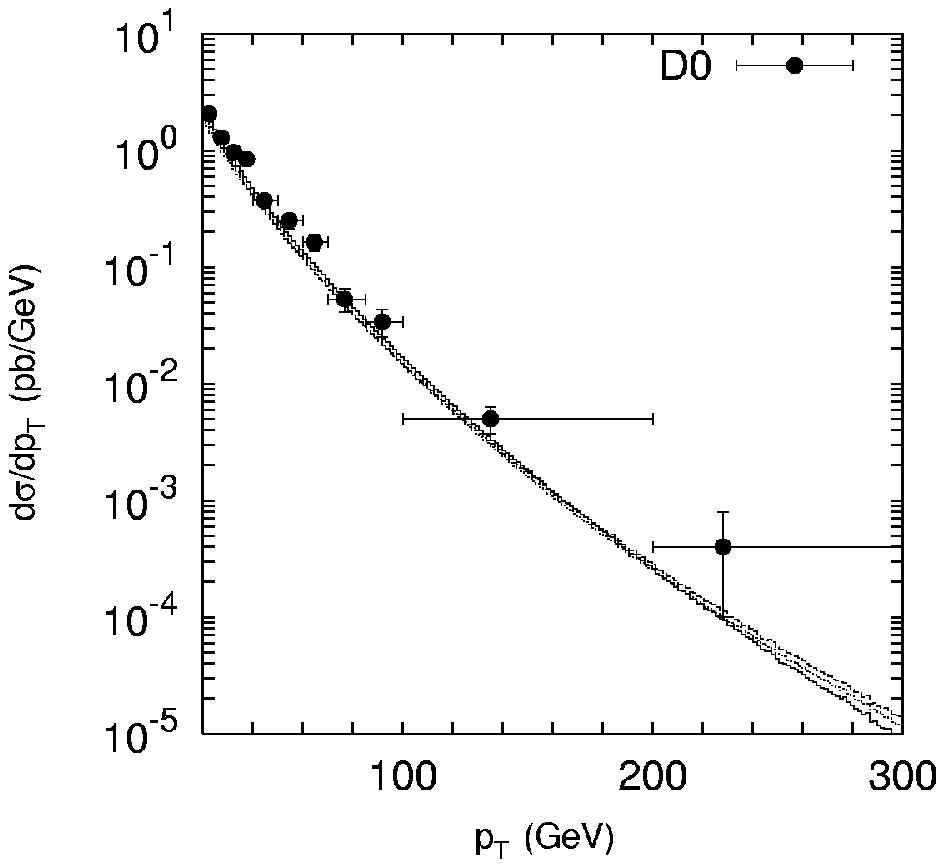, width = 18cm}
\caption{Transverse mometum distribution of
the $Z^0$ boson production. 
Notations of histograms are the same as in Fig.~7.
The cross sections times branching fraction $f(Z \to l^+l^-)$ are shown.
The experimental data are from D$\oslash$~[8].}
\end{center}
\label{fig11}
\end{figure}

\newpage

\begin{figure}
\begin{center}
\epsfig{figure=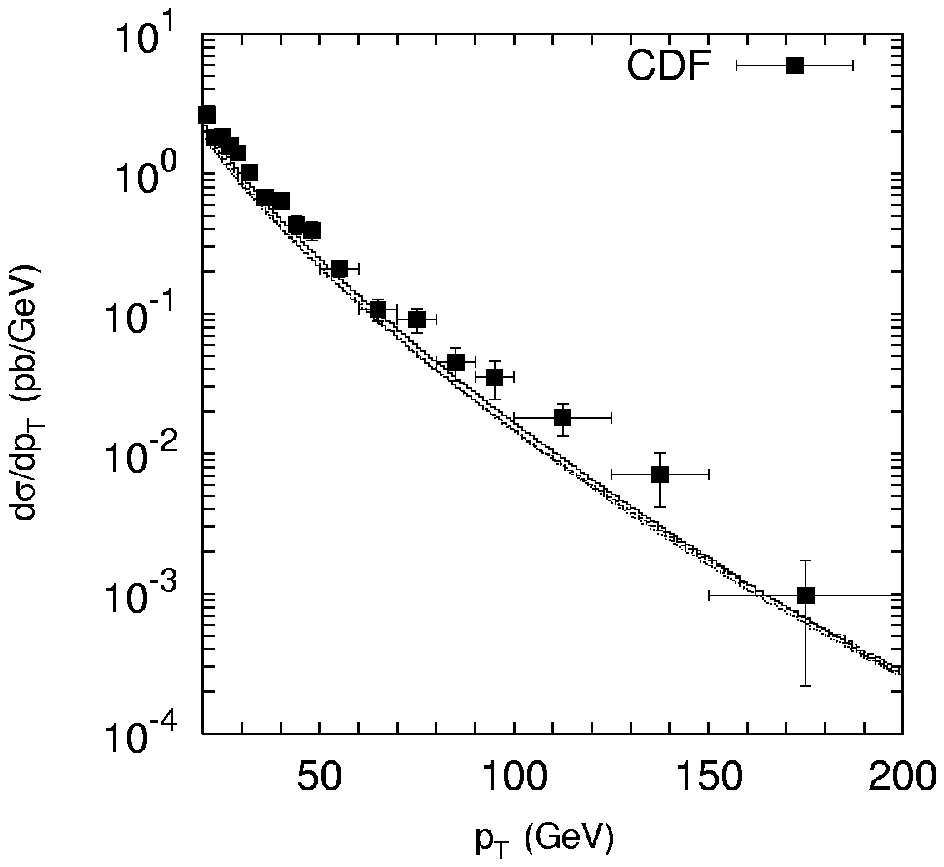, width = 18cm}
\caption{Transverse mometum distribution of
the $Z^0$ boson production. 
Notations of histograms are the same as in Fig.~7.
The cross sections times branching fraction $f(Z \to l^+l^-)$ are shown.
The experimental data are from CDF~[4].}
\end{center}
\label{fig12}
\end{figure}

\newpage

\begin{figure}
\begin{center}
\epsfig{figure=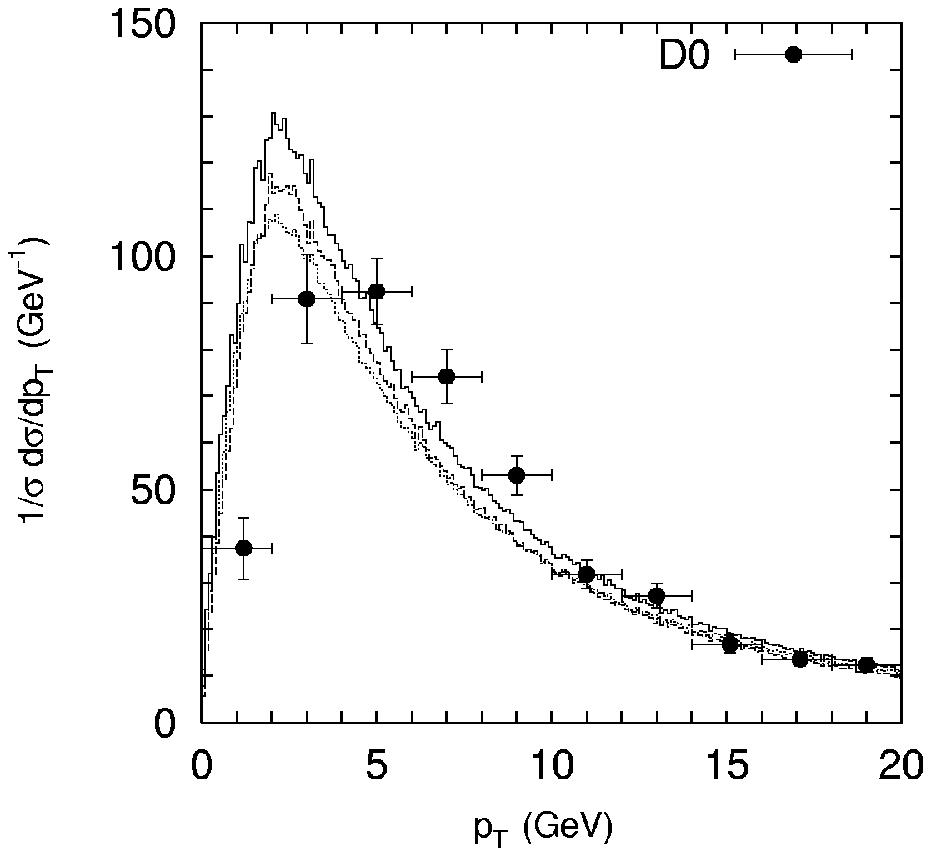, width = 18cm}
\caption{Normalized transverse mometum distribution of
the $W^\pm$ boson production. Notations of histograms are 
the same as in Fig.~7. 
The earlier experimental data are from D$\oslash$~[5].}
\end{center}
\label{fig13}
\end{figure}

\newpage

\begin{figure}
\begin{center}
\epsfig{figure=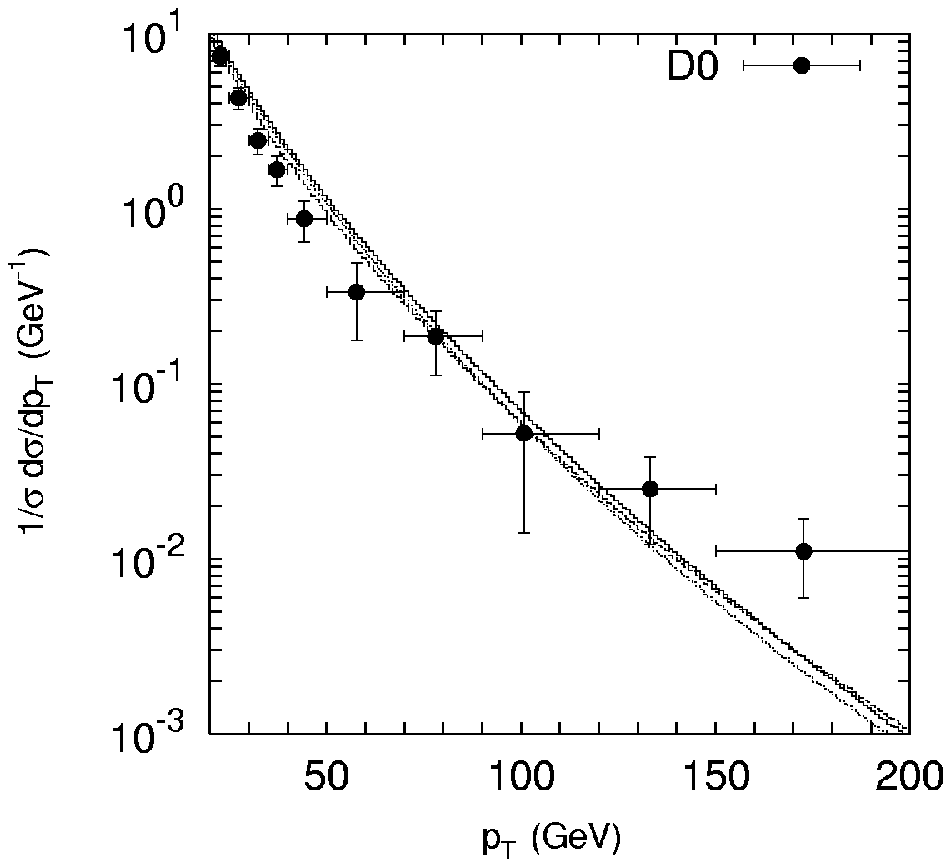, width = 18cm}
\caption{Normalized transverse mometum distribution of
the $W^\pm$ boson production. Notations of histograms are 
the same as in Fig.~7. 
The earlier experimental data are from D$\oslash$~[5].}
\end{center}
\label{fig14}
\end{figure}

\newpage

\begin{figure}
\begin{center}
\epsfig{figure=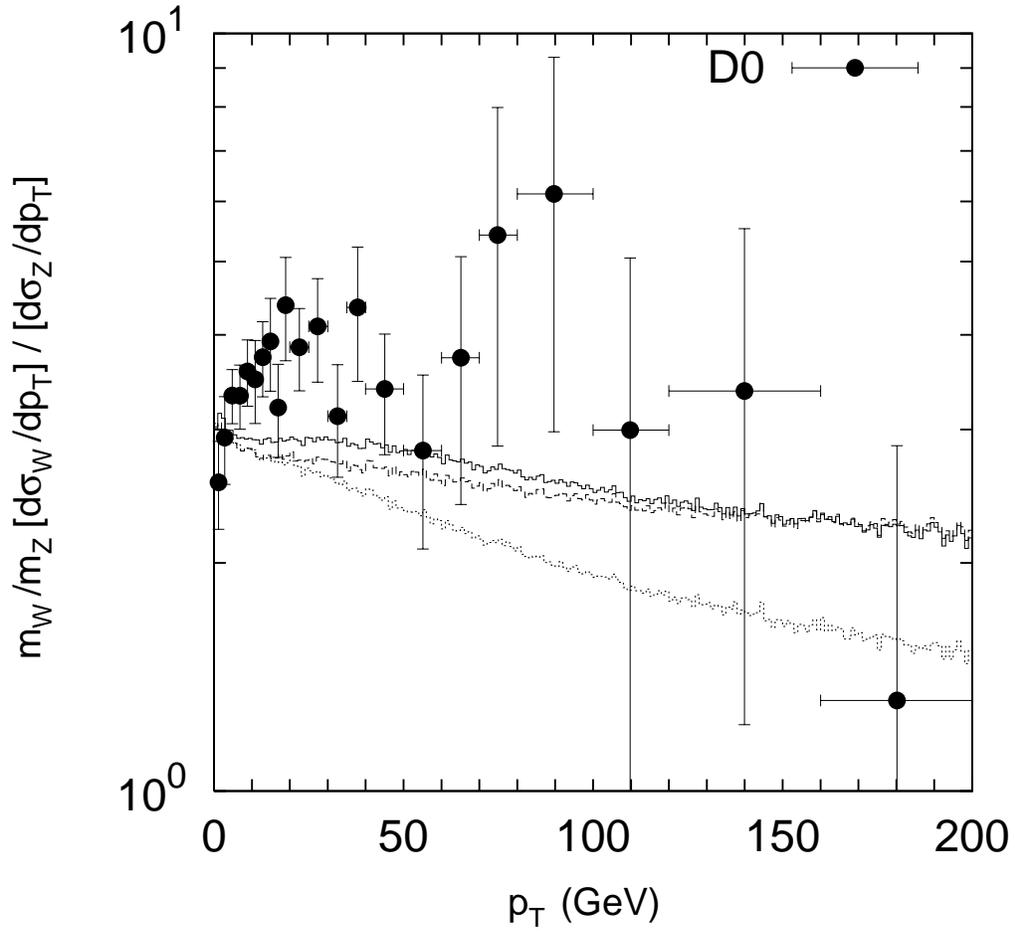, width = 18cm}
\caption{Ratio of differential cross section for $W^\pm$ to $Z^0$ 
production. Notations of histograms are 
the same as in Fig.~7. 
The experimental data are from D$\oslash$~[8].}
\end{center}
\label{fig15}
\end{figure}

\newpage

\begin{figure}
\begin{center}
\epsfig{figure=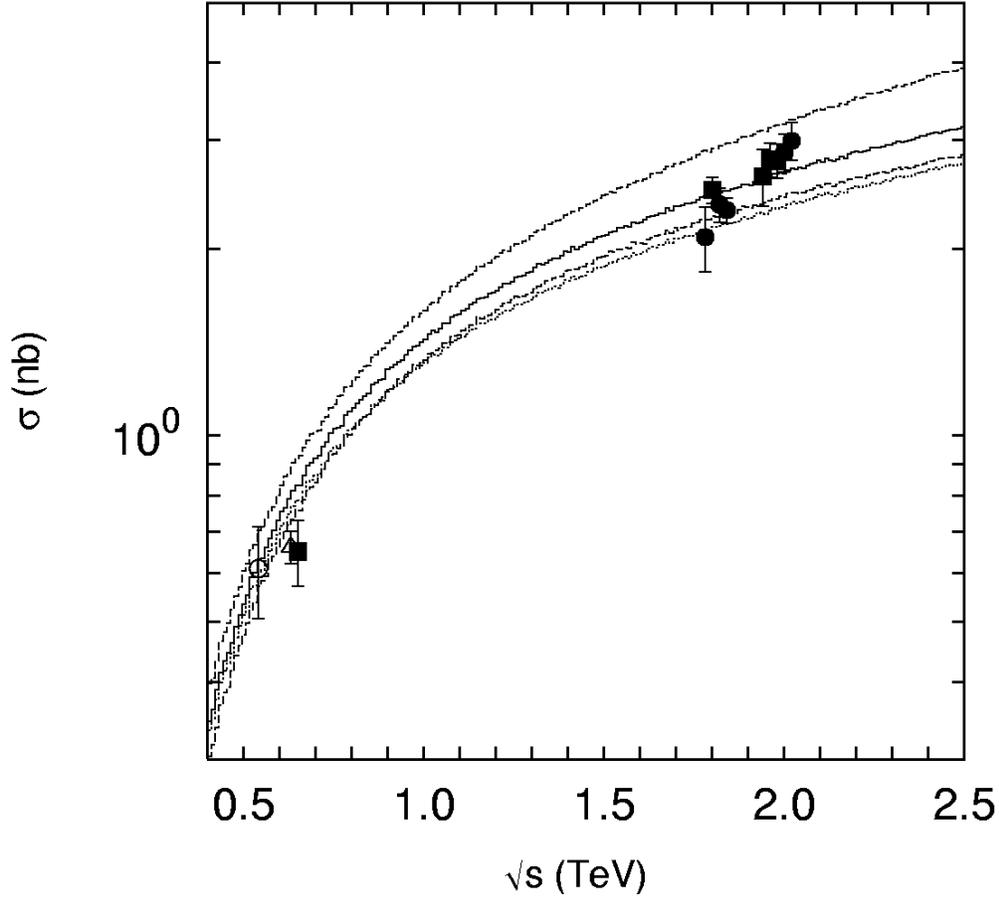, width = 18cm}
\caption{The total cross section of the inclusive $W^\pm$ boson production
as a function of $\sqrt s$. The solid and dotted histograms 
correspond to the results obtained with the CCFM 
and KMR unintegrated gluon densities, respectively. The upper and 
lower dashed histograms 
correspond to the scale variations in CCFM gluon density 
as it was described in text.
The cross sections times branching fraction $f(W \to l\nu)$ are shown.
The experimental data are from UA1~[1], UA2[2], D$\oslash$~[6, 7] and CDF~[3].}
\end{center}
\label{fig16}
\end{figure}

\newpage

\begin{figure}
\begin{center}
\epsfig{figure=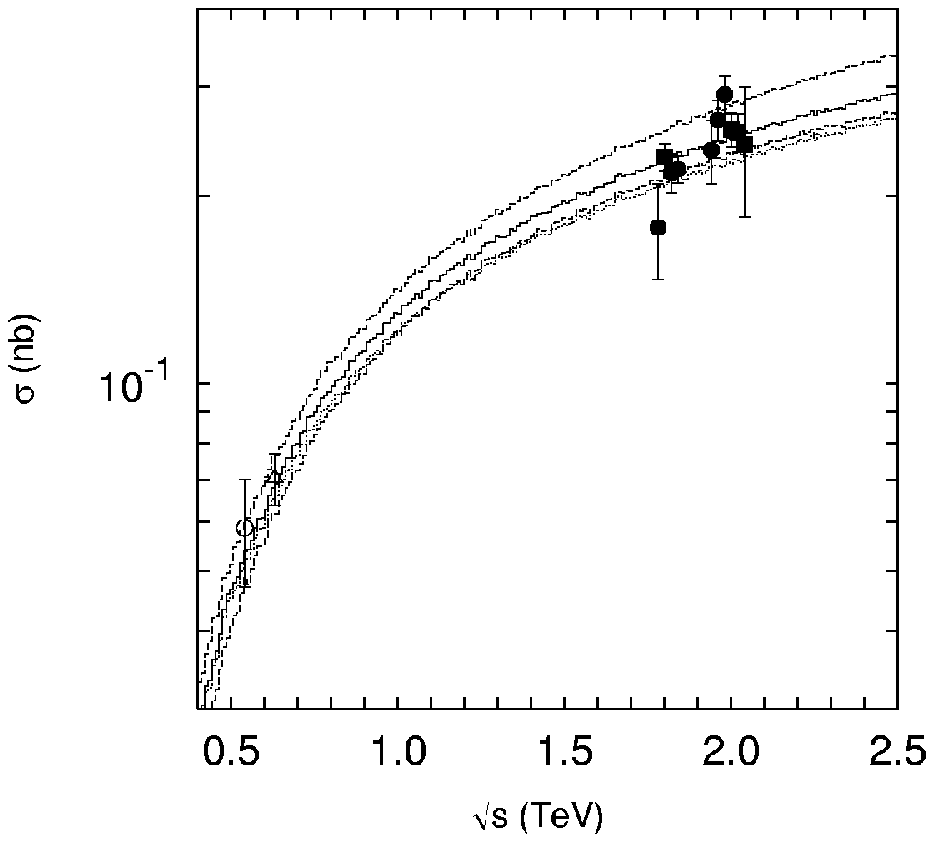, width = 18cm}
\caption{The total cross section of the inclusive $Z^0$ boson production
as a function of $\sqrt s$.
Notations of histograms are 
the same as in Fig.~16.
The cross sections times branching fraction $f(Z \to l^+l^-)$ are shown.
The experimental data are from UA1~[1], UA2[2], D$\oslash$~[6, 7] and CDF~[3].}
\end{center}
\label{fig17}
\end{figure}


\begin{thebibliography}{41}

\bibitem{1} C.~Albajar {\sl et al.} (UA1 Collaboration), Phys. Lett. B {\bf 253}, 503 (1991).
\bibitem{2} J.~Alitti {\sl et al.} (UA2 Collaboration), Z. Phys. {\bf 47}, 11 (1990).
\bibitem{3} F.~Abe {\sl et al.} (CDF Collaboration), Phys. Rev. Lett. {\bf 76}, 3070 (1996).
\bibitem{4} B.~Affolder {\sl et al.} (CDF Collaboration), Phys. Rev. Lett. {\bf 84}, 845 (2000).
\bibitem{5} B.~Abbott {\sl et al.} (D$\oslash$ Collaboration), Phys. Rev. Lett. {\bf 80}, 5498 (1998).
\bibitem{6} S.~Abachi {\sl et al.} (D$\oslash$ Collaboration), Phys. Rev. Lett. {\bf 75}, 1456 (1995).
\bibitem{7} B.~Abbott {\sl et al.} (D$\oslash$ Collaboration), Phys. Rev. D {\bf 61}, 072001 (2000).
\bibitem{8} B.~Abbott {\sl et al.} (D$\oslash$ Collaboration), Phys. Rev. D {\bf 61}, 032004 (2000).
\bibitem{9} B.~Abbott {\sl et al.} (D$\oslash$ Collaboration), Phys. Lett. B {\bf 513}, 292 (2001).
\bibitem{10} P.~Sutton, A.~Martin, R.~Roberts, and W.J.~Stirling, Phys. Rev. D {\bf 45}, 2349 (1992).
\bibitem{11} R.~Rijken and W.~van~Neerven, Phys. Rev. D {\bf 51}, 44 (1995).
\bibitem{12} R.~Hamberg, W.~van~Neerven, and T.~Matsuura, Nucl. Phys. B {\bf 359}, 343 (1991).
\bibitem{13} R.~Harlander and W.~Kilgore, Phys. Rev. Lett. {\bf 88}, 201801 (2002).
\bibitem{14} W.~van~Neerven and E.~Zijstra, Nucl. Phys. B {\bf 382}, 11 (1992).
\bibitem{15} J.~Collins, D.~Soper, and G.~Sterman, Nucl. Phys. B {\bf 250}, 199 (1985);\\
  J.~Collins and D.~Soper, Nucl. Phys. B {\bf 193}, 381 (1981); Nucl. Phys. B {\bf 197}, 446 (1982). 
\bibitem{16} C.~Davies, B.~Webber, and W.J.~Stirling, Nucl. Phys. B {\bf 256}, 413 (1985);\\
  C.~Davies and W.J.~Stirling, Nucl. Phys. B {\bf 244}, 337 (1984).
\bibitem{17} G.~Altarelli, R.K.~Ellis, M.~Grego, and G.~Martinelli, Nucl. Phys. B {\bf 246}, 12 (1984).
\bibitem{18} P.B.~Arnold and R.~Kauffman, Nucl. Phys. B {\bf 349}, 381 (1991).
\bibitem{19} G.A.~Ladinsky and C.P.~Yuan, Phys. Rev. D {\bf 50}, 4239 (1994).
\bibitem{20} R.K.~Ellis and S.~Veseli, Nucl. Phys. B {\bf 511}, 649 (1998).
\bibitem{21} C.~Balazs and C.P.~Yuan, Phys. Rev. D {\bf 56}, 5558 (1997).
\bibitem{22} A.~Kulesza and W.J.~Stirling, Nucl. Phys. B {\bf 555}, 279 (1999).
\bibitem{23} L.V.~Gribov, E.M.~Levin, and M.G.~Ryskin, Phys. Rep. {\bf 100}, 1 (1983);\\
  E.M.~Levin, M.G.~Ryskin, Yu.M.~Shabelsky and A.G.~Shuvaev, Sov. J. Nucl. Phys. {\bf 53}, 657 (1991).
\bibitem{24} S.~Catani, M.~Ciafoloni and F.~Hautmann, Nucl. Phys. B {\bf 366}, 135 (1991);\\
  J.C.~Collins and R.K.~Ellis, Nucl. Phys. B {\bf 360}, 3 (1991).
\bibitem{25} E.A.~Kuraev, L.N.~Lipatov, and V.S.~Fadin, Sov. Phys. JETP {\bf 44}, 443 (1976);\\
  E.A.~Kuraev, L.N.~Lipatov, and V.S.~Fadin, Sov. Phys. JETP {\bf 45}, 199 (1977);\\
  I.I.~Balitsky and L.N.~Lipatov, Sov. J. Nucl. Phys. {\bf 28}, 822 (1978).
\bibitem{26} M.~Ciafaloni, Nucl. Phys. B {\bf 296}, 49 (1988);\\
  S.~Catani, F.~Fiorani, and G.~Marchesini, Phys. Lett. B {\bf 234}, 339 (1990);\\
  S.~Catani, F.~Fiorani, and G.~Marchesini, Nucl. Phys. B {\bf 336}, 18 (1990);\\
  G.~Marchesini, Nucl. Phys. B {\bf 445}, 49 (1995).
\bibitem{27} V.N.~Gribov and L.N.~Lipatov, Yad. Fiz. {\bf 15}, 781 (1972);\\
  L.N.~Lipatov, Sov. J. Nucl. Phys. {\bf 20}, 94 (1975);\\
  G.~Altarelli and G.~Parisi, Nucl. Phys. B {\bf 126}, 298 (1977);\\
  Y.L.~Dokshitzer, Sov. Phys. JETP {\bf 46}, 641 (1977).
\bibitem{28} A.~Gawron and J.~Kwiecinski, Phys. Rev. D {\bf 70}, 014003 (2004).
\bibitem{29} B.~Andersson {\sl et al.} (Small-$x$ Collaboration), Eur. Phys. J. C {\bf 25}, 77 (2002);\\
  J.~Andersen {\sl et al.} (Small-$x$ Collaboration), Eur. Phys. J. C {\bf 35}, 77 (2004);\\
  J.~Andersen {\sl et al.} (Small-$x$ Collaboration), Eur. Phys. J. C {\bf 48}, 53 (2006).
\bibitem{30} G.~Watt, A.D.~Martin, and M.G.~Ryskin, Phys. Rev. D {\bf 70}, 014012 (2004).
\bibitem{31} M.A.~Kimber, A.D.~Martin, and M.G.~Ryskin, Phys. Rev. D {\bf 63}, 114027 (2001);\\
  G.~Watt, A.D.~Martin, and M.G.~Ryskin, Eur. Phys. J. C {\bf 31}, 73 (2003).
\bibitem{32} A.V.~Bogdan and A.V.~Grabovsky, Nucl. Phys. B {\bf 773}, 65 (2007).
\bibitem{33} S.P.~Baranov, A.V.~Lipatov and N.P~Zotov, Phys. Rev. D {\bf 77}, 074024 (2008).
\bibitem{34} J.A.M.~Vermaseren, "Symbolic Manipulation with FORM", published by Computer
  Algebra Nederland, Kruislaan 413, 1098, SJ Amsterdaam, 1991; ISBN 90-74116-01-9.
\bibitem{35} R.E.~Prange, Phys. Rev. {\bf 110}, 240 (1958);\\
  S.P.~Baranov, Phys. Atom. Nucl. {\bf 60}, 1322 (1997).
\bibitem{36} G.P.~Lepage, J. Comput. Phys. {\bf 27}, 192 (1978).  
\bibitem{37} M.~Gl\"uck, E.~Reya and A.~Vogt, Phys. Rev. D {\bf 46}, 1973 (1992);\\
  M.~Gl\"uck, E.~Reya and A.~Vogt, Z. Phys. C {\bf 67}, 433 (1995). 
\bibitem{38} H.~Jung, A.V.~Kotikov, A.V.~Lipatov and N.P.~Zotov, in Proceedings 
  of ICHEP'06, p.~493 [hep-ph/0611093].
\bibitem{39} H.~Jung, Mod. Phys. Lett. A {\bf 19}, 1 (2004).
\bibitem{40} W.-M.~Yao {\sl et al.} (Particle Data Group), J. Phys. G {\bf 33}, 1 (2006).
\bibitem{41} M.~Deak and F.~Schwennsen, arXiv:0805.3763 [hep-ph].

\end{thebibliography}
\end{document}